\documentclass[usenatbib]{mnras}

\usepackage[T1]{fontenc}
\usepackage{easyReview}
\usepackage[normalem]{ulem}
\usepackage{cancel}

\DeclareRobustCommand{\VAN}[3]{#2}
\let\VANthebibliography\thebibliography
\def\thebibliography{\DeclareRobustCommand{\VAN}[3]{##3}\VANthebibliography}

\pdfoutput=1 
\usepackage{amsmath}	
\usepackage{amssymb}	
\usepackage{color}
\usepackage{txfonts}
\usepackage{natbib}
\usepackage{pdflscape} 
\usepackage{graphicx}	
\usepackage{subcaption} 
\usepackage{booktabs}   
\usepackage{hyperref}  
\usepackage{tabularx}  
\usepackage{ragged2e}    
\usepackage{siunitx}     
\usepackage{multirow}    
\usepackage{threeparttable} 
\usepackage{bm} 
\title[Photometric zero-points for the SDSS system]{Recalibration of SDSS photometric zero-points based on the InfraRed Flux Method temperature scale}

\author
[Z. H. Zhou et al.]{
Zenghua Zhou,$^{1,2,4,6}$\thanks{{E-mail:zhouzenghua@ynao.ac.cn}}
Luca Casagrande,$^{2}$
Heran Xiong,$^{2}$
Yanjun Guo,$^{1,4,6}$
\newauthor%
{Jiajia Li, $^{5,6}$
Zhanwen Han,$^{1,3,4,6}$
and Xuefei Chen$^{1,3,4,6}$}\thanks{E-mail: cxf@ynao.ac.cn},
\\
$^{1}$Yunnan Astronomical Observatories, Chinese Academy of Sciences, Kunming 650011, China; \href{zhouzenghua@ynao.ac.cn} \href{cxf@ynao.ac.cn}\\
$^{2}$Research School of Astronomy and Astrophysics, Australian National University, Weston Creek ACT 2611, Australia\\
$^{3}$Key Laboratory for the Structure and Evolution of Celestial Objects, Chinese Academy of Sciences, Kunming 650011, China\\
$^{4}$International Centre of Supernovae, Yunnan Key Laboratory, Kunming 650216, China\\
$^{5}$National Astronomical Observatories, Chinese Academy of Sciences, Beijing, 100012, China\\
$^{6}$University of Chinese Academy of Sciences, Beijing 100049, China\\
}
\date{Accepted ; Received ; in original form}

\pubyear{2026}

\newcommand{\teff}{{T_{\rm eff}}}

\begin{document}
\label{firstpage}
\pagerange{\pageref{firstpage}--\pageref{lastpage}}
\maketitle

\begin{abstract}
Accurate photometric zero-points are essential for translating observed magnitudes into physical fluxes, from comparing with models to ensuring consistency across surveys. We determine the zero-points needed to place the Sloan Digital Sky Survey (SDSS) $ugriz$ system on its nominal AB definition, by exploiting the sensitivity of the Infrared Flux Method (IRFM) to broadband flux calibration. Using benchmark effective temperatures for over 6,000 FGK-type stars, we invert the method to identify the zero-point corrections required for SDSS photometry to reproduce the adopted temperature scale.

The $r$ band is found to be very well standardized, while the $i$ and $z$ bands show offsets of a few hundredths of a magnitude, consistent with previous studies. We also find a small offset in the $g$ band. The largest discrepancy occurs in the $u$ band, where the derived offset depends strongly on the adopted filter transmission curves, in particular whether one uses the original definition commonly adopted in the literature or the updated measurements that account for the presence of a red leak. This effect introduces a colour-dependent zero-point offset that becomes apparent when using a sample of late-type stars. Independent comparisons with CALSPEC spectrophotometric standards and Gaia XP spectra broadly support the offsets derived from the IRFM analysis.

Our results provide a revised set of SDSS zero-points anchored to the IRFM temperature scale and demonstrate that large stellar samples can be used to constrain photometric calibration. The methodology presented here offers a complementary approach to traditional spectrophotometric calibration and may prove useful for future large-scale surveys.
\end{abstract}




\section{Introduction}

Modern astrophysics increasingly relies on large photometric surveys to derive physical properties of stars and galaxies, compare observations across instruments, and confront theoretical models with data. Achieving these goals requires photometric systems to be placed on a well-defined absolute flux scale, such that measured magnitudes can be reliably translated into physical fluxes and compared across surveys. Even small zero-point offsets can propagate into systematic errors in stellar parameters, distances, and population studies, underscoring the importance of accurate photometric standardisation.

This need is particularly acute in the context of current and upcoming wide-field surveys, including the Sloan Digital Sky Survey \citep[SDSS;][]{York2000AJ}, the Legacy Survey of Space and Time \citep[LSST;][]{LSST2019ApJ}, the Chinese Space Station Telescope \citep[CSST;][]{ZhanHu2011SSPMA,CaoYe2018MNRAS,Gong2019ApJ}, and the Nancy Grace Roman Space Telescope \citep{Roman2020}. The scientific exploitation of these datasets critically depends on robust cross-survey calibration and internally consistent photometric zero-points.

SDSS represents a landmark project in modern astronomy, providing homogeneous photometry in five broadband filters ($ugriz$) spanning approximately 300--1100\,nm \citep{Fukugita1996AJ,Doi2010AJ}. Its magnitude system is explicitly tied to the AB scale \citep{Oke1983ApJ}, enabling a direct connection between observed magnitudes and physical flux densities, while the use of asinh magnitudes \citep{Lupton1999AJ} ensures robust measurements at low signal-to-noise ratios.

In practice, however, the realised SDSS photometric system departs subtly from its idealised AB definition owing to instrumental effects, atmospheric variability, and complexities in the calibration pipeline. These factors introduce small but non-negligible zero-point offsets. Numerous studies have attempted to quantify these corrections, often reaching inconsistent conclusions. For example, comparisons with HST spectrophotometric standards suggest an offset of about $-0.04$ mag in the $u$ band (i.e. $u_{AB}$ = $u_{SDSS} - 0.04$), while $g$, $r$, and $i$ are generally consistent with AB \citep{Bohlin2001AJ,Holberg2006AJ}, and the $z$ band shows evidence of an offset around $+0.02$ \citep{SDSS_DR2,Holberg2006AJ}. Further refinements by \citet{Eisenstein2006ApJS} proposed $z_{AB}$ = $z_{SDSS} + 0.03$, and a minor i-band adjustment $i_{AB}$ = $i_{SDSS} + 0.015$. Other analyses instead argue for internal consistency at the $\sim$0.01\,mag level \citep{BBSuzuki2018AJ}. Resolving these discrepancies and establishing a coherent set of SDSS zero-points remains essential.

A promising approach to this problem exploits the strong connection between broadband photometry and stellar effective temperatures. Among the various techniques for determining $\teff$, 
the Infrared Flux Method (IRFM) offers a direct link to the fundamental definition of $\teff$ as it was originally developed to achieve high accuracy measurements of stellar angular diameters \citep{Blackwell1977MNRAS,Blackwell1979MNRAS,Blackwell1980A&A, Blackwell1994A&A}. 

Because the IRFM relies primarily on broadband photometry and requires minimal theoretical input, it provides a powerful, largely model-independent temperature scale. Crucially, the IRFM depends on how magnitudes are converted into physical fluxes. This sensitivity can be inverted: if reliable reference effective temperatures are available, photometric zero-points can be inferred. This methodology was introduced by \citet{Casagrande2019MNRAS} and successfully applied to the SkyMapper photometric system.

In this work, we extend that framework to SDSS. Using benchmark effective temperatures from \citet{Casagrande2021MNRAS}, which implement Gaia $BP$ and $RP$ photometry within the IRFM and are calibrated on an absolute scale using solar twins and interferometry \citep{Casagrande2010A&A,Casagrande2021MNRAS}, we determine the SDSS $ugriz$ zero-points by quantifying how different zero-point choices affect the IRFM temperature scale. 

\begin{figure}
    \centering
    \includegraphics[width=\columnwidth]{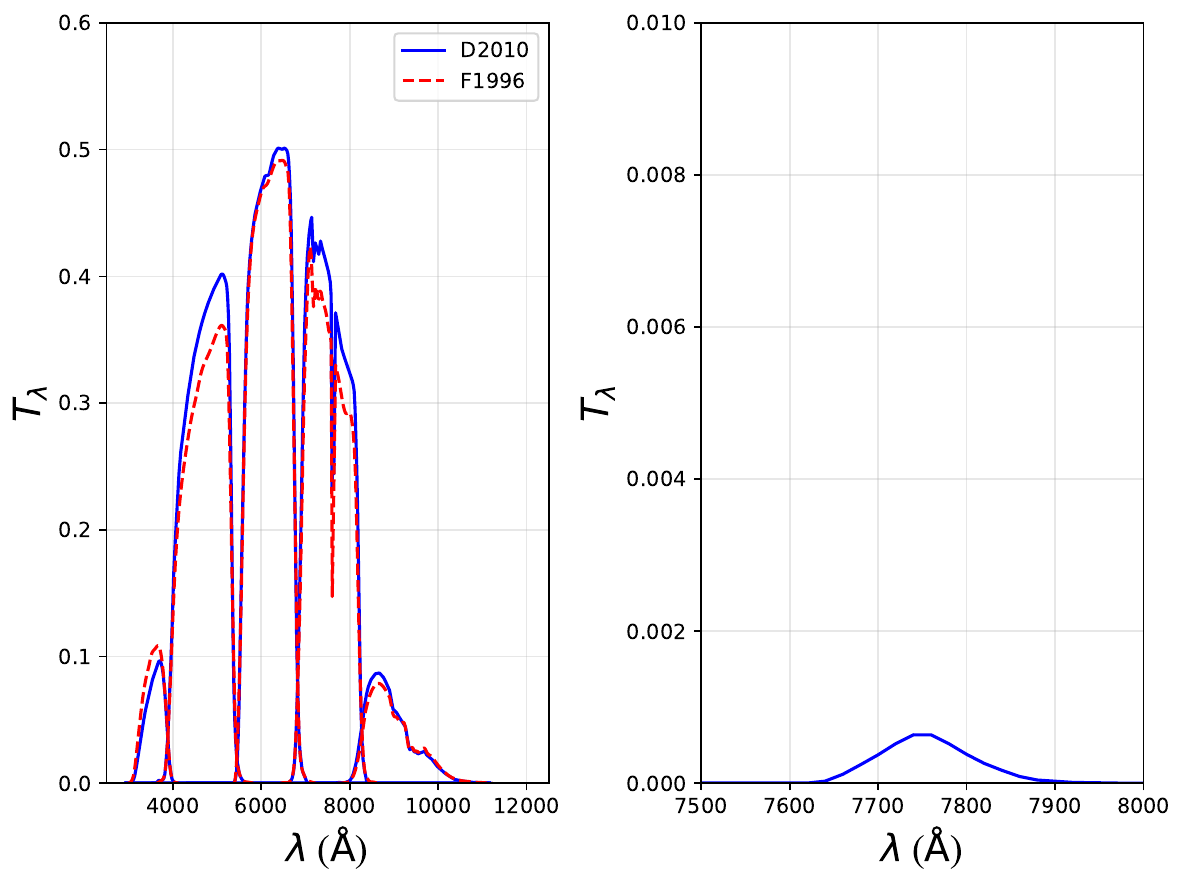}
    \caption{
        Total system throughput of the SDSS \textit{ugriz} photometric filters. 
        \textbf{Left}: The original transmission curves from \citet[][red dashed lines]{Fukugita1996AJ} and the updated curves from \citet[][blue solid lines]{Doi2010AJ}. 
        \textbf{Right}: Red-leak contribution in the \textit{u}-band reported in \protect\citet{Doi2010AJ}, illustrating residual sensitivity at $\lambda > 760$ nm. 
        }
    \label{fig:SDSS_throughput}
\end{figure}

The paper is organized as follows. Section~\ref{sec:Meth} describes the methodology and data used to determine the SDSS zero-points. Section~\ref{sec:results} presents the derived $ugriz$ zero-points along with their uncertainties. In Section~\ref{sec:discus}, we test and validate our results against independent approaches, with summary and conclusions in Section~\ref{sec:concl}. 

\section{Methodology}\label{sec:Meth}

\subsection{The impact of photometric zero-points on the IRFM}\label{sec:IRFM_AB}

A source with flux $f_{\lambda}$ observed in a filter $\zeta$ characterized by a system response $T_{\zeta}$ over the wavelength range $\lambda_i$ to $\lambda_f$ has an AB magnitude
\begin{equation}\label{eq:magABlam}
m_{\zeta,AB} = -2.5 \log \frac{\int_{\lambda_i}^{\lambda_f} \lambda f_{\lambda} T_\zeta \,d\lambda}{f_\nu^0 c \int_{\lambda_i}^{\lambda_f} \frac{T_\zeta}{\lambda}\,d\lambda},
\end{equation}
where $f_\nu^0$ is the constant flux density per unit frequency defining zero AB magnitude and $c$ is the speed of light \citep[e.g.,][]{Casagrande2014MNRAS}. 

In real observing systems ($m_\zeta$), departures from this idealized definition are unavoidable, and band-dependent zero-point offsets $\epsilon_\zeta'$ are therefore commonly present:
\begin{equation}\label{eq:SDSSvsAB}
m_\zeta = m_{\zeta,AB} + \epsilon_\zeta'.
\end{equation}
Such offsets arise from a combination of instrumental throughput variations and calibration effects, and propagate directly into inferred physical fluxes as well as comparisons between observations and model predictions. Since in literature zero-points are usually defined as those needed to standardize to a given system (AB in the case of this work), in the rest of the paper we adhere to this definition\footnote{We note for completeness that this definition of $\epsilon_\zeta$ is opposite to the one adopted in Table 1 of \cite{Casagrande2014MNRAS}.}, i.e. $\epsilon_\zeta = -\epsilon_\zeta'$.

Our approach exploits the sensitivity of the Infrared Flux Method (IRFM) to photometric zero-points. Because the IRFM derives effective temperatures from broadband flux ratios, known reference temperatures can be used to invert the problem and solve for the zero-points that reproduce a target $\teff$ scale. We apply this framework to the SDSS photometric system, following the methodology introduced by \citet{Casagrande2019MNRAS}.

Briefly, the IRFM determines stellar effective temperatures by comparing the ratio of bolometric to infrared monochromatic flux measured at Earth with the corresponding ratio defined at the stellar surface. Working in the infrared minimizes sensitivity to stellar parameters such as metallicity and surface gravity \citep[e.g.,][]{Blackwell1991A&A,Alonso1996A&AS,Casagrande2006MNRAS}, thereby reducing the problem to an accurate determination of stellar fluxes from photometry. In our implementation, these fluxes are obtained using SDSS optical bands together with 2MASS infrared photometry, following the prescriptions in \citet{Casagrande2010A&A,Casagrande2019MNRAS}. While the infrared component of the IRFM is always anchored to 2MASS, different optical systems (SDSS $ugriz$ or Gaia $BP/RP$ in this work) are adopted as required. Thus, in the following, references to a given optical system being implemented in the IRFM implicitly assume the simultaneous use of 2MASS photometry in the infrared.

When effective temperatures are fixed, the procedure can be inverted by varying optical photometric zero-points until the IRFM reproduces a reference $\teff$ scale. In practice, owing to the magnitude range of the SDSS system, we derive $ugriz$ zero-points by requiring that $\teff$ values obtained by implementing SDSS photometry in the IRFM lie on the same scale as those derived for the same stars using Gaia $BP$ and $RP$ photometry within the IRFM \citep{Casagrande2021MNRAS}. The temperature scale in that work is calibrated in an absolute sense using solar twins, with its accuracy independently confirmed by both solar twins and interferometry \citep{Casagrande2010A&A,Casagrande2021MNRAS}. The reference sample used here is further described in Section \ref{sec:data}.

Zero-point solutions are obtained using two sets of SDSS response functions (Fig.~\ref{fig:SDSS_throughput}): the original throughput from \citet[][hereafter F96]{Fukugita1996AJ} and the updated measurements from \citet[][hereafter D10]{Doi2010AJ}. The main difference between these two sets regards the red-leak in $u$ band, which is responsible for a colour-dependent zero-point offset as we discuss in what follows. Our analysis focuses on transmission curves from D10, which are a more realistic representation of the SDSS filters used to derive published $ugriz$ magnitudes. We provide results for F96 in Appendix, along with a brief discussion of the differences between the two. 

\subsection{Zero-point derivation}\label{sec:IRFM_SDSS}


To determine the photometric zero-points of the SDSS optical bands, we implement one filter $\zeta$ at a time within the IRFM (with the understanding that 2MASS photometry is always used in the infrared). For each band we explore $T_{\rm eff,\zeta}^\epsilon$, i.e., how the resulting effective temperatures vary across a wide range of trial zero-points $\epsilon_\zeta$. 

Because each star has a different fraction of its bolometric flux encompassed by a given band, depending primarily on its effective temperature, a change in zero-point affects the derived effective temperatures differently along the temperature sequence. Ignoring this behaviour would introduce an artificial colour dependence in the derived zero-points. To account for this effect, we explicitly model how the IRFM temperature responds to a change in the zero-point as a function of effective temperature. For each SDSS band we compute the temperature difference:
\begin{equation}
\Delta T_{\rm eff,\zeta}^{\epsilon} = T_{\rm eff,\zeta}^{\epsilon} - T_{\rm eff,\zeta}^{0},
\end{equation}
where $T_{\rm eff,\zeta}^{0}$ is obtained adopting SDSS photometry with no zero-point corrections, and $T_{\rm eff,\zeta}^{\epsilon}$ corresponds to the IRFM temperature after applying a reference zero-point shift. The dependence of this difference on effective temperature is then described by fitting a fourth-order polynomial $P(T_{\rm eff,\zeta}^{0})$, which accurately captures the trend across the stellar sample.

By comparing IRFM solutions computed for different values of $\epsilon_\zeta$, we verify empirically that the temperature response scales linearly with the applied zero-point shift.
The ratio of the resulting temperature differences confirms the expected proportionality, allowing the polynomial response $P$ to be scaled when sampling different values of $\epsilon_\zeta$.

We use this polynomial to build a data-driven model whereby the zero-point dependent effective temperature of the $i$-th star can be written as:
\begin{equation}\label{eq:model}
T_{{\rm eff},\zeta,i}^{\rm model} (\epsilon_\zeta) = \frac{\epsilon_\zeta}{s_{\zeta}} \, P(T_{{\rm eff},\zeta,i}^0) + T_{{\rm eff},\zeta,i}^0
\end{equation}
where $s_{\zeta}$ is the empirically determined scaling factor relating the polynomial response to a $0.01$ mag zero-point change. Given a sample of N benchmark stars, the optimal zero-point for each band is obtained by minimizing:
\begin{equation}\label{eq:chi2}
\chi^2(\epsilon_\zeta) =
\sum_{i=1}^{N}
\left[
\frac{T_{{\rm eff},\zeta,i}^{\rm model}(\epsilon_\zeta) - T_{{\rm eff},i}^{\rm ref}}{\sigma_i}
\right]^2 ,
\end{equation}
where 
$\sigma_i$ represents the adopted uncertainty in the temperature differences, and $T_{{\rm eff},i}^{\rm ref}$ are the benchmark effective temperatures described in the next Section. 
The best-fitting $\epsilon_\zeta$ corresponds to the minimum $\chi^2$, while the $1\sigma$ uncertainty is obtained from the range of $\epsilon_\zeta$ for which $\chi^2=\chi^2_{\rm min}+1$.

This procedure is performed independently for each of the five SDSS passbands, yielding the set of zero-point corrections required for the SDSS photometry to reproduce the adopted reference temperature scale.

\subsection{Reference sample}\label{sec:data}

Our analysis requires a sample of stars with accurately determined effective temperatures ($T_{\rm{eff}}^{\rm ref}$) to serve as benchmarks for the IRFM inversion. These reference temperatures are used to anchor the SDSS-based IRFM solutions and thereby determine the SDSS photometric zero-points. The only requirement on the stellar sample is the availability of high-quality SDSS, Gaia, and 2MASS photometry, together with homogeneous stellar parameters.

For the southern hemisphere, we adopt the sample of \citet{Casagrande2021MNRAS}, who applied the IRFM using Gaia DR3 photometry to stars with metallicities and surface gravities from the GALAH DR3 spectroscopic sample. For consistency in the northern hemisphere, we apply the same Gaia-based IRFM to APOGEE stars whose stellar parameters have been placed onto the GALAH DR3 scale using \textsc{The Cannon} \citep{Govind2022MNRAS}, thereby ensuring a homogeneous set of atmospheric parameters across both hemispheres.

Each star is cross-matched with SDSS DR13 for $ugriz$, Gaia DR3 for $BP$ and $RP$, and 2MASS for $JHK_s$ magnitudes. Reddening corrections are adopted following \citet{Casagrande2021MNRAS}, based on either the maps of \citet{Green2019ApJ} or \citet{sfd}. Crucially, for any given star the same reddening values are applied when implementing either Gaia or SDSS photometry in the IRFM. To further minimize systematic effects, we restrict our sample to stars with low reddening and well-determined spectroscopic parameters (see below). Finally, it is worth emphasizing that our analysis is differential: consequently, the precise choice of [Fe/H], $\log g$, or extinction map has only a secondary impact, provided the same values are used consistently when implementing different photometric systems.

The initial cross-match yields several tens of thousands of stars. From these, we select a high-quality subsample by requiring:
\begin{itemize}
\item reliable spectroscopic parameters as reported in Table \ref{tab:quality_cuts};
\item best Gaia DR3 $BP$ and $RP$ photometry and standard photometric processing pipeline used (\texttt{phot\_proc\_mode == 0});
\item photometric uncertainties $<0.04$\,mag in all SDSS bands and $\leq0.05$\,mag in 2MASS;
\item Good SDSS bitmasks via CasJob\footnote{https://skyserver.sdss.org/dr13/en/help/cooking/general/pdf/flags.pdf};
\item clean photometry and stellar classification flags: \texttt{clean == 1} (indicating no artifacts, saturation, or blending) and \texttt{type == 6} (restricting to point sources/stars).
\end{itemize}

\begin{figure*}
    \centering
        \includegraphics[width=0.33\textwidth]{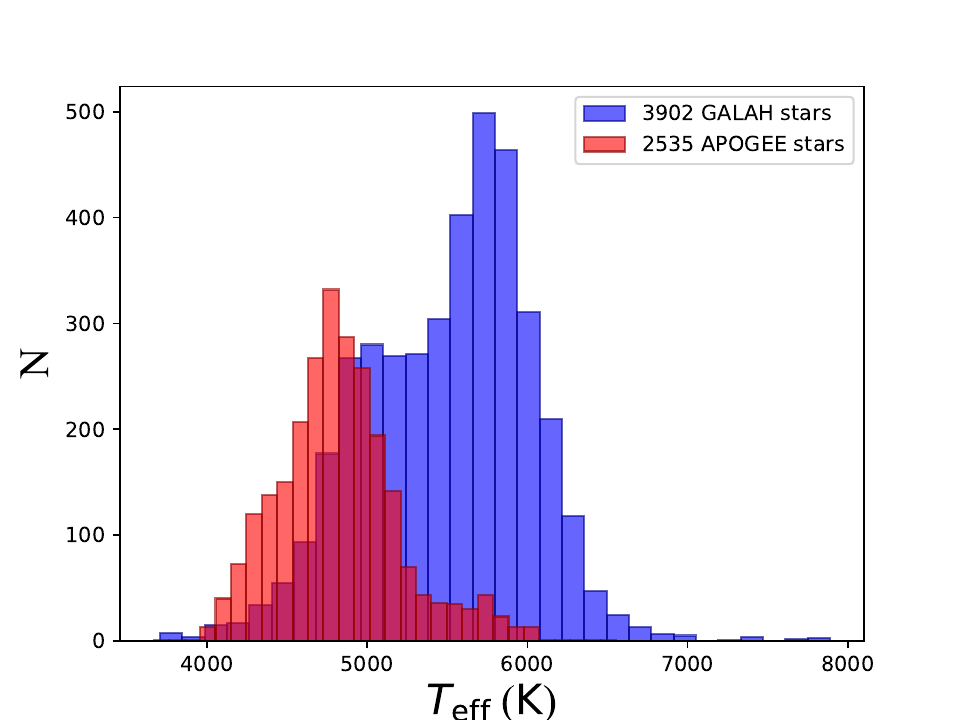}
        \includegraphics[width=0.33\textwidth]{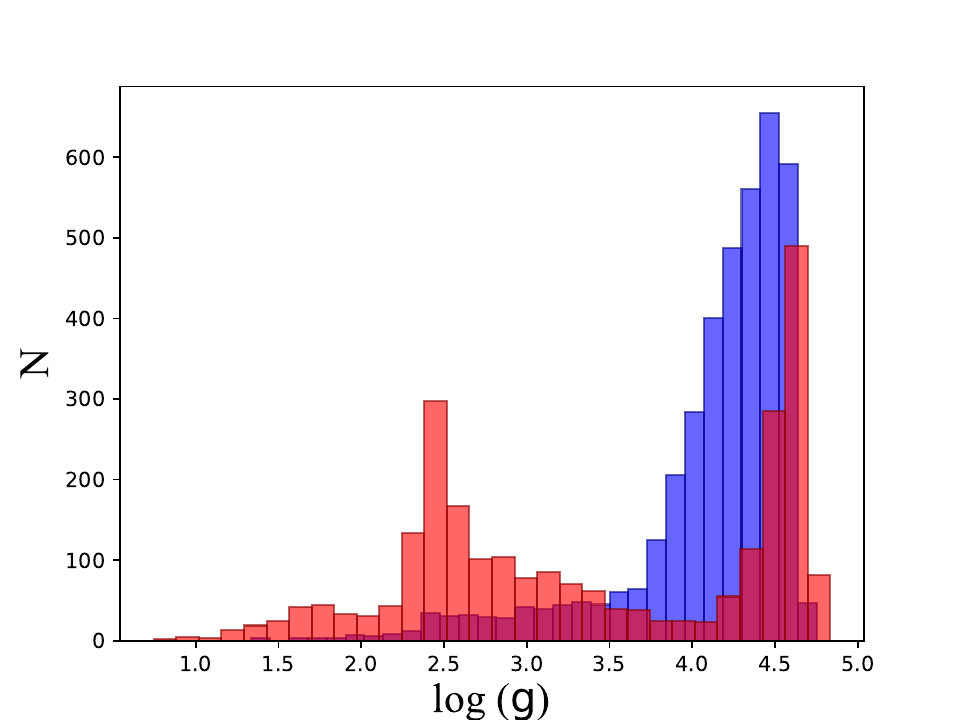}
        \includegraphics[width=0.33\textwidth]{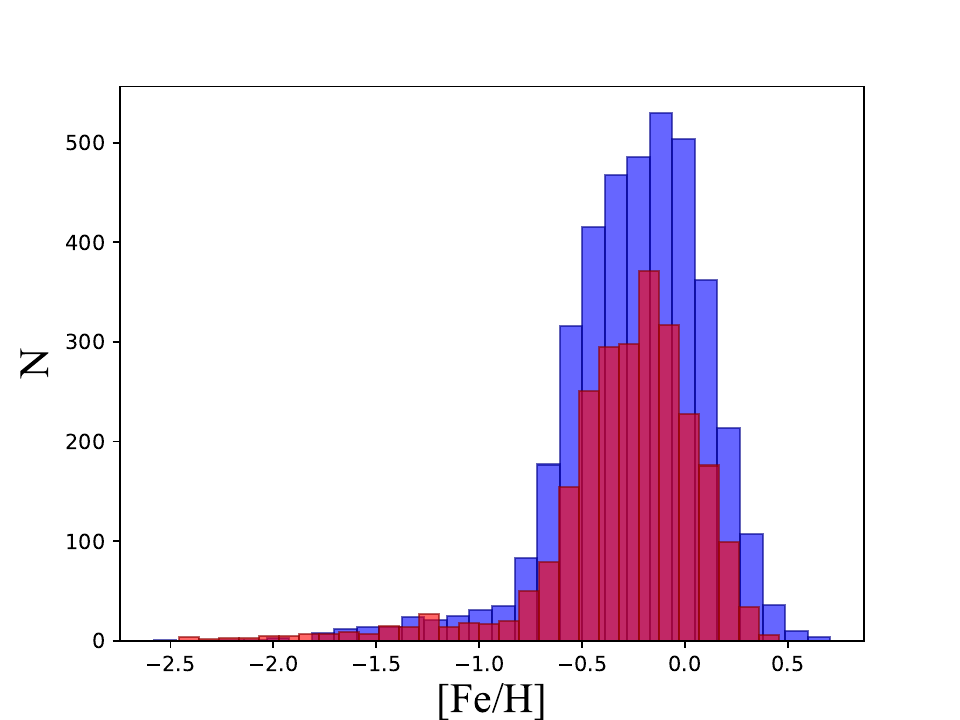}
    \caption{Distribution of stellar parameters for our GALAH and APOGEE samples.}
    \label{fig:sample_data}
\end{figure*}

After applying these criteria, the final reference sample consists of 3902 GALAH stars and 2535 APOGEE stars, spanning a wide range in effective temperature, metallicity, and surface gravity (Fig.~\ref{fig:sample_data}).


\begin{table}
    \centering
    \caption{Spectroscopic quality flags for stars in the GALAH and APOGEE catalogues. In all instances a further cut on spectroscopic uncertainties $\sigma(\log g)<0.2$\,dex, $\sigma([\mathrm{Fe/H}])<0.1$\,dex and spectroscopic $\sigma(T_{\rm eff})<160$\,K has been applied.}
    \label{tab:quality_cuts}
    \small  
    \begin{tabularx}{\linewidth}{@{} l >{\RaggedRight}X @{}}
        \toprule
        \textbf{Criterion} & \textbf{Description} \\
        \midrule
        \multicolumn{2}{l}{\textbf{GALAH DR3 Spectroscopic quality cuts}} \\
        \midrule
        \texttt{flag\_sp == 0} & Good GALAH stellar parameters (no issues in spectroscopic analysis) \\
        \midrule
        \multicolumn{2}{l}{\textbf{APOGEE Spectroscopic quality cuts (Nandamukar et al. 2022)}} \\
        \midrule
        \texttt{flag\_id == 1} & APOGEE spectra \\
        \texttt{flag\_survey != 1} & Exclude invalid survey pipeline estimates  \\
        \texttt{flag\_spectra == 0} & No issues with spectra \\
        \texttt{flag\_Cannon\_dist == 0} & Only stars with Cannon parameters within the training set parameters \\
        \bottomrule
    \end{tabularx}
\end{table}

\section{Results} \label{sec:results}

\begin{figure*}
    \includegraphics[scale=0.8]{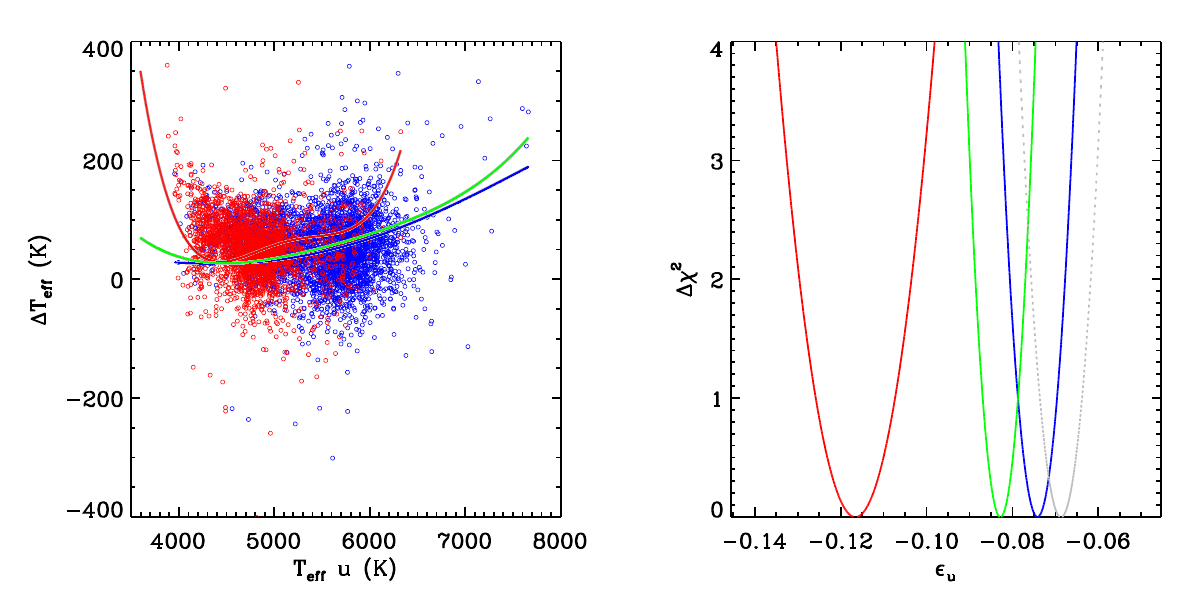}
    \caption{Left panel: $\Delta T_{\rm{eff}} = T_{{\rm eff}}^{\rm ref} - T_{{\rm eff},u}(\epsilon_u=0)$ for stars in the GALAH (blue) and APOGEE (red) samples. Continuous lines show the best fits for the GALAH (blue), APOGEE (red) and combined (green) samples. Right panels: $\Delta \chi^2 = \chi^2 - \chi_{\rm min}^2$ curves, whereby values of $1$, $2$ and $3$ define sigma confidence levels. Colours refer to the same samples shown in the left panel. The gray line corresponds to the combined sample restricted to stars with temperatures $>5500$~K.}
    \label{fig:epsilon_u}
\end{figure*}

\begin{figure*}
    \includegraphics[scale=0.8]{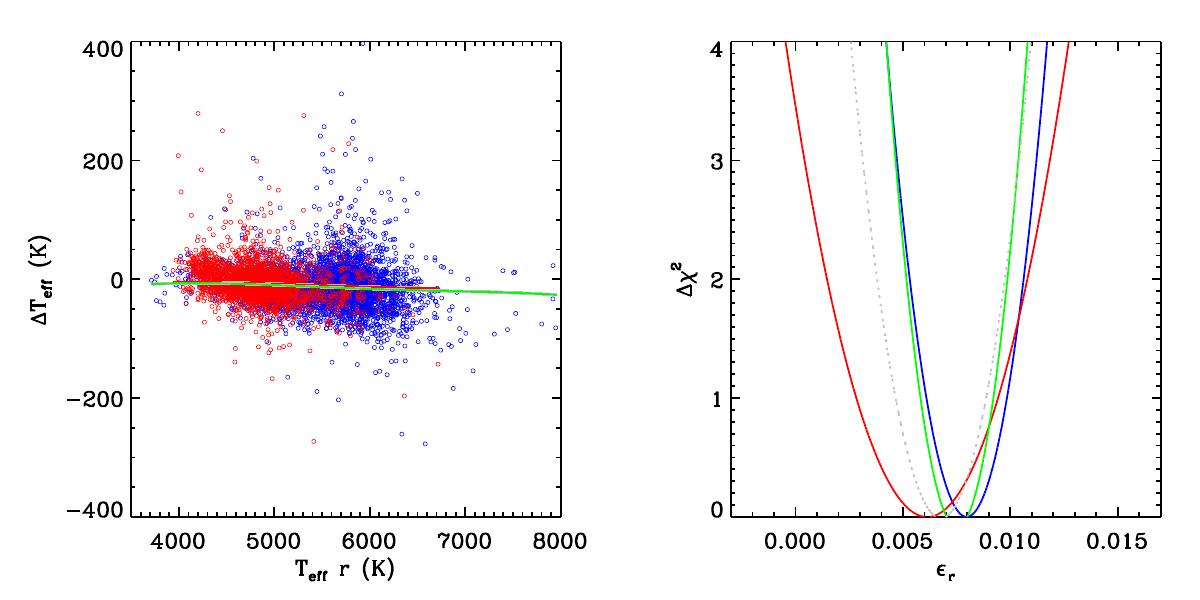}    
    \caption{Same as Figure \ref{fig:epsilon_u}, but for the $r$ band.}
    \label{fig:epsilon_r}
\end{figure*}

For all stars in our GALAH and APOGEE samples, the left panels of Figures~\ref{fig:epsilon_u} and \ref{fig:epsilon_r} show the comparison between our reference effective temperatures and those obtained by implementing a single SDSS band with no zero-point correction in the IRFM. It is immediately apparent that while in the $r$ band the temperature difference is nearly a constant offset, this is not the case for the $u$ band. This behaviour reflects the varying contribution that each band makes to the bolometric flux for stars with different stellar parameters, and is well described by the $\epsilon_\zeta$-dependent empirical model of Equation~\ref{eq:model}. We sample a broad range of possible $\epsilon_\zeta$ values and determine the best fit to the data by minimizing the $\chi^2$ defined in Equation~\ref{eq:chi2}. The best-fitting solutions are shown as continuous lines, while the right panels display the corresponding $\Delta\chi^2 = \chi^2 - \chi^2_{\rm min}$ curves for the GALAH, APOGEE, and combined samples.

For all bands, the formal significance levels are at the few-millimagnitude level, reflecting the intrinsic precision of the IRFM inversion when applied to samples of several thousand stars. However, realistic uncertainties must also account for systematic effects, in particular the impact of shifts in the zero-point of the effective temperature scale, which propagate directly into the inferred photometric zero-points. The uncertainty on the zero-point of our adopted $\teff$ scale is of order $20$~K \citep{Casagrande2010A&A,Casagrande2021MNRAS}. This translates into shifts in the derived photometric zero-points ranging from a few hundredths of a magnitude in the $u$ band to about $0.01$ mag in the $z$ band. Accounting for this systematic contribution is therefore essential in order to derive realistic uncertainties. The resulting zero-points for the $ugriz$ bands are summarized in Table~\ref{tab:zp}.

\begin{table}
	\centering
	\caption{SDSS zero-points $\epsilon_\zeta$ derived through the IRFM analysis. Error bars account for a chance of $\pm20$~K in the zero-point of the adopted reference $\teff$ scale.}
	\label{tab:zp}
	\begin{tabular}{lrrr} 
		\hline
		    & GALAH & APOGEE & COMBINED\\
		\hline
		$u$ & $-0.074 \pm 0.026$ & $-0.117 \pm 0.043$ & $-0.083 \pm0.030$\\
		$g$ &  $0.027 \pm 0.018$ & $ 0.018 \pm 0.025$ & $ 0.024 \pm0.019$ \\
		$r$ &  $0.008 \pm 0.011$ & $ 0.006 \pm 0.016$ & $ 0.008 \pm0.012$\\
		$i$ &  $0.034 \pm 0.009$ & $ 0.047 \pm 0.013$ & $ 0.037 \pm0.010$\\
		$z$ &  $0.046 \pm 0.008$ & $ 0.061 \pm 0.011$ & $ 0.050 \pm0.009$\\        
		\hline
	\end{tabular}
\end{table}

The most striking feature is the much larger offset in the $u$ band compared to the values typically quoted in the literature from early SDSS calibration studies, which generally lie in the range $-0.03$ to $-0.04$ mag, as summarized in the Introduction. In the Appendix we show that this difference arises from the adoption of the D10 response curves: using the original F96 curves yields values consistent with those previously published. This behaviour is linked to the presence of the red leak, which introduces a colour-dependent zero-point offset.

The APOGEE sample contains cooler stars on average than the GALAH sample, which explains the larger offsets derived from APOGEE. If both samples are restricted to stars with effective temperatures above $5500$~K, the $u$-band zero-points shift to $-0.067$ and $-0.092$ for the GALAH and APOGEE samples, respectively. The only other band affected at more than the $0.01$ mag level by such a restriction is the $z$ band for the APOGEE sample.

Since the APOGEE and GALAH samples probe different hemispheres, another possibility would be a systematic zero-point offset in the adopted reference $\teff$ scale, which is based on Gaia and 2MASS photometry. We consider this unlikely, as modifying the $\teff$ scale to bring the $u$-band results into agreement would significantly worsen the consistency between the GALAH and APOGEE samples in the $griz$ bands. 

We therefore interpret the discrepancy in the $u$ band as evidence for a colour-dependent zero-point offset. This conclusion is further strengthened by the comparison with CALSPEC standards presented in the next section.

Aside from the larger offset in the $u$ band, our analysis reveals significant offsets in the $g$, $i$, and $z$ bands, while the $r$ band appears to be very well standardized. The offset we derive for the $i$ band is larger than that reported by \citet{Eisenstein2006ApJS}, while our result for the $z$ band is broadly consistent with both the sign and magnitude found in that work.

In the next section we use the CALSPEC library together with Gaia XP spectra to investigate these offsets in greater detail.

\section{Validation and Discussion} \label{sec:discus}

\subsection{CALSPEC Standards}

The CALSPEC library represents the gold standard for absolute spectrophotometry, providing flux calibrations traceable to laboratory standards with a precision of order $\sim1\%$ in absolute flux \citep{Bohlin2014AJ,Bohlin2022stis,Bohlin2025AJ}.

Here we use CALSPEC absolute fluxes ($f_{\lambda}$) together with the D10 system response functions ($T_\zeta$) to compute AB magnitudes ($m_{\zeta,AB}$) via Equation~(\ref{eq:magABlam}). Comparison with the observed SDSS magnitudes then allows the determination of zero-point offsets ($\epsilon_{\zeta}$) through Equation~(\ref{eq:SDSSvsAB}).

\begin{figure}
    \centering
    \includegraphics[scale=0.42]{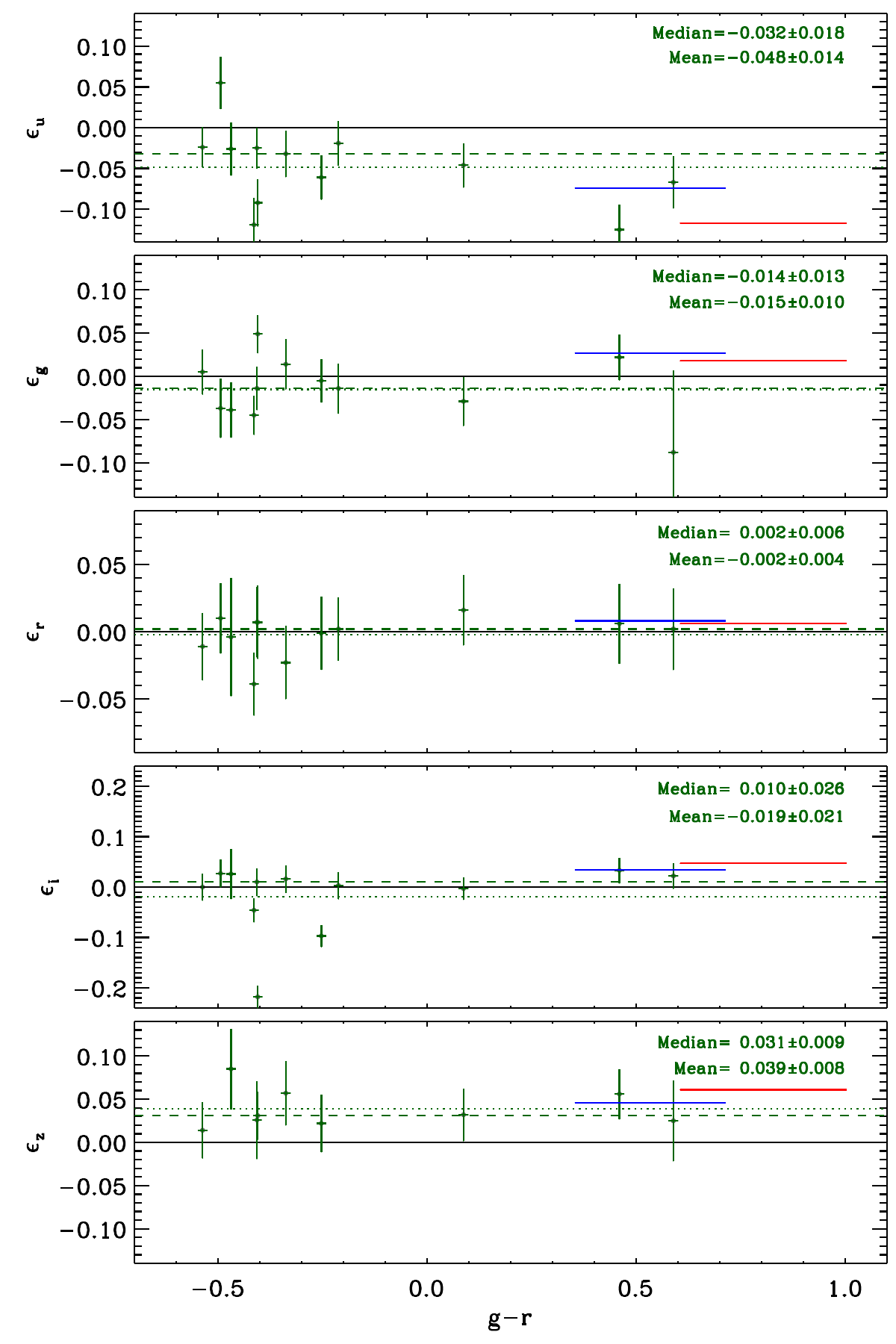}
    \caption{Crosses show the difference between AB magnitudes computed from CALSPEC fluxes and observed SDSS magnitudes, $\epsilon_\zeta = m_{\zeta,AB}-m_{\zeta,SDSS}$, as a function of observed $g-r$ colour (green). Error bars are obtained by adding in quadrature the observational and CALSPEC uncertainties. Dashed and dotted lines indicate the median and mean offsets, respectively, reported in each panel together with their associated standard deviations. The blue and red horizontal bars show the zero-points derived from the GALAH and APOGEE samples (Table~\ref{tab:zp}), with their lengths corresponding to the 16th and 84th percentiles of the colour distributions of the samples.}
    \label{fig:calspec_D10}
\end{figure}

Cross-matching the CALSPEC database (as of October 2024) with the SDSS DR13 catalogue yields 48 stars in common. Applying stringent photometric quality criteria (\texttt{clean == 1} and \texttt{type == 6}) reduces this to 14 stars. From this sample, we exclude vB8 and 2M0036+18, which have spectral types M7\,V and L3.5\,V, corresponding to $\teff \sim 2500$~K and $\sim 2000$~K, respectively \citep[e.g.,][]{c08,dahn02}, significantly well below the range explored in this work\footnote{For 2M0036+18 CALSPEC flux is only available longward of 529~nm. While such cool stars often exhibit significant variability, magnitude differences are consistent with offsets shown in Figure \ref{fig:calspec_D10}, apart for $\epsilon_u \sim 1.5$ mag for vB8.}. For another 3 stars we did not consider $z$ magnitudes since their CALSPEC fluxes stop at 1020nm, slightly shorter of the 1100nm limit of this filter. Photometric uncertainties are computed by combining the statistical and systematic flux errors provided in the CALSPEC files for each star.

Figure~\ref{fig:calspec_D10} shows that most CALSPEC standards have $g-r < 0$, although a few overlap with the colour range spanned by the GALAH and APOGEE samples. Notably, the two CALSPEC stars with $g-r \simeq 0.5$ exhibit $u$ band offsets that encompass the range of values derived from the IRFM analysis. In contrast, Figure~A3 shows that this trend is weaker when using the older F96 transmission curves, which do not account for the red leak and therefore underestimate its impact, particularly for cooler stars.

Our zero-point determinations for the $r$ and $z$ bands are in very good agreement with the CALSPEC results in both sign and value. The only band showing a noticeable discrepancy is the $g$ band, where the offsets derived from the two approaches differ. This is investigated further through the use of XP spectra. 

\subsection{Validation with Gaia DR3 Spectrophotometry}\label{GaiaXP}

We further validate our findings using Gaia XP spectrophotometry, which covers the wavelength range 336--1020\,nm. This spectrophotometry is calibrated using a network of standards ultimately tied to the CALSPEC flux scale \citep{pancino}. Since the stars in our sample are relatively bright, we restrict our analysis to XP sampled spectra, although using the continuous spectra yields very similar results. The flux calibration accuracy of Gaia XP spectrophotometry at the few percent level \citep[e.g.,][]{Montegriffo2023A&A} makes it well suited for an independent assessment of photometric zero-points. 

Because Gaia photometry was used to construct our benchmark $\teff$ sample, which itself is tied to the absolute temperature scale of \citet{Casagrande2010A&A}, this comparison provides both an independent validation and a useful consistency check of the overall calibration framework.

Following the same procedure described in the previous Section, we compute synthetic SDSS magnitudes from the XP spectra and compare them with the observed ones to derive $\epsilon_\zeta$. Because of the wavelength coverage of the XP spectra, as well as residual features at the bluest and reddest wavelengths \citep[e.g.,][]{Montegriffo2023A&A}, we restrict this comparison to the $gri$ filters. For these bands the differences in the F96 transmission curves are negligible, resulting in virtually identical offsets. 

\begin{figure}
    \centering
    \includegraphics[scale=0.42]{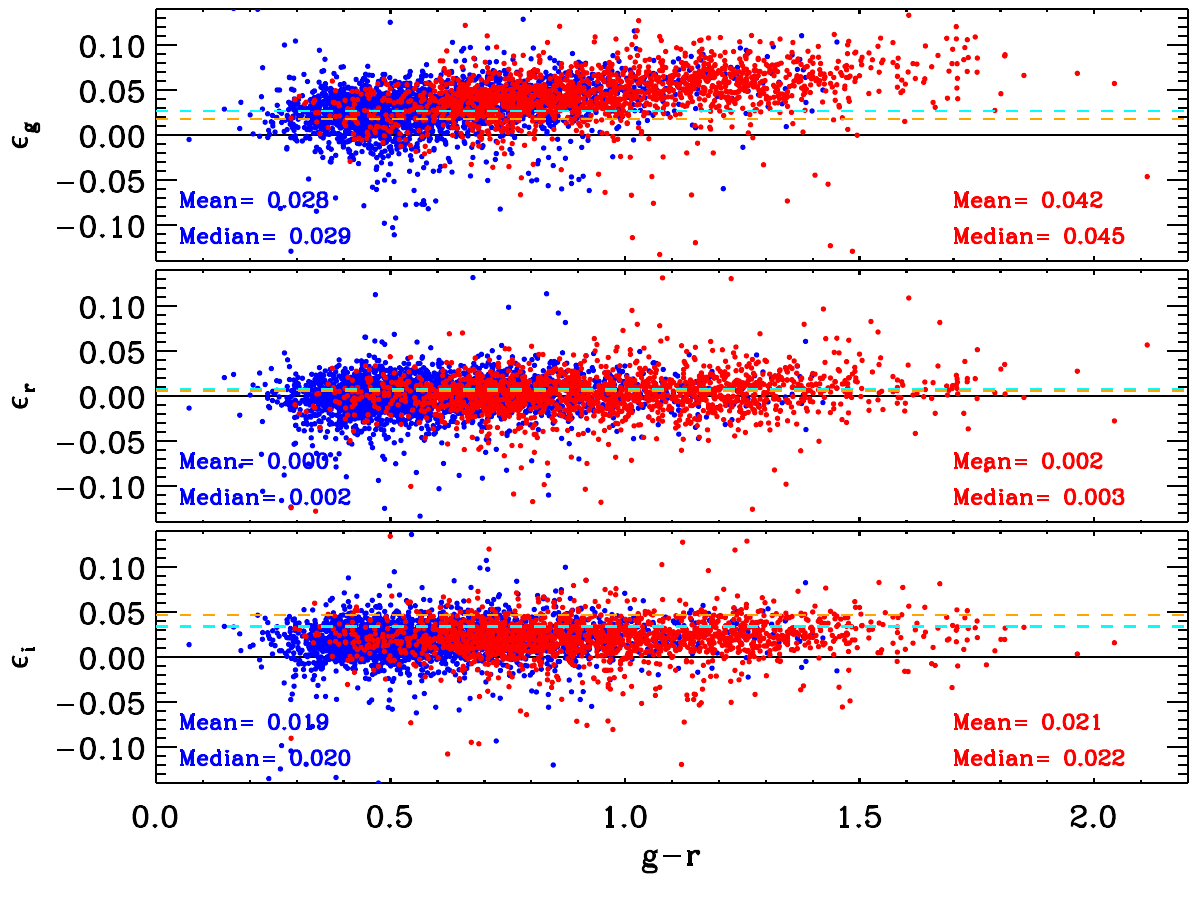}
    \caption{Same as Figure \ref{fig:calspec_D10} using Gaia XP spectrophotometry for the GALAH (blue) and APOGEE (red) samples. Dashed lines are offsets from IRFM inversion of the same stars from Table \ref{tab:zp} (cyan for GALAH, orange for APOGEE).}
    \label{fig:xp_D10}
\end{figure}

Overall, the zero-points derived from the IRFM inversion broadly agree in both sign and magnitude with those inferred from Gaia XP spectra. There is excellent agreement for the $r$ band, confirming the high level of standardization of this band already indicated by the IRFM inversion and the CALSPEC comparison. For the $g$ and $i$ bands we confirm the presence of significant offsets inferred from the IRFM analysis. 

\begin{figure}
    \centering
    \includegraphics[scale=0.42]{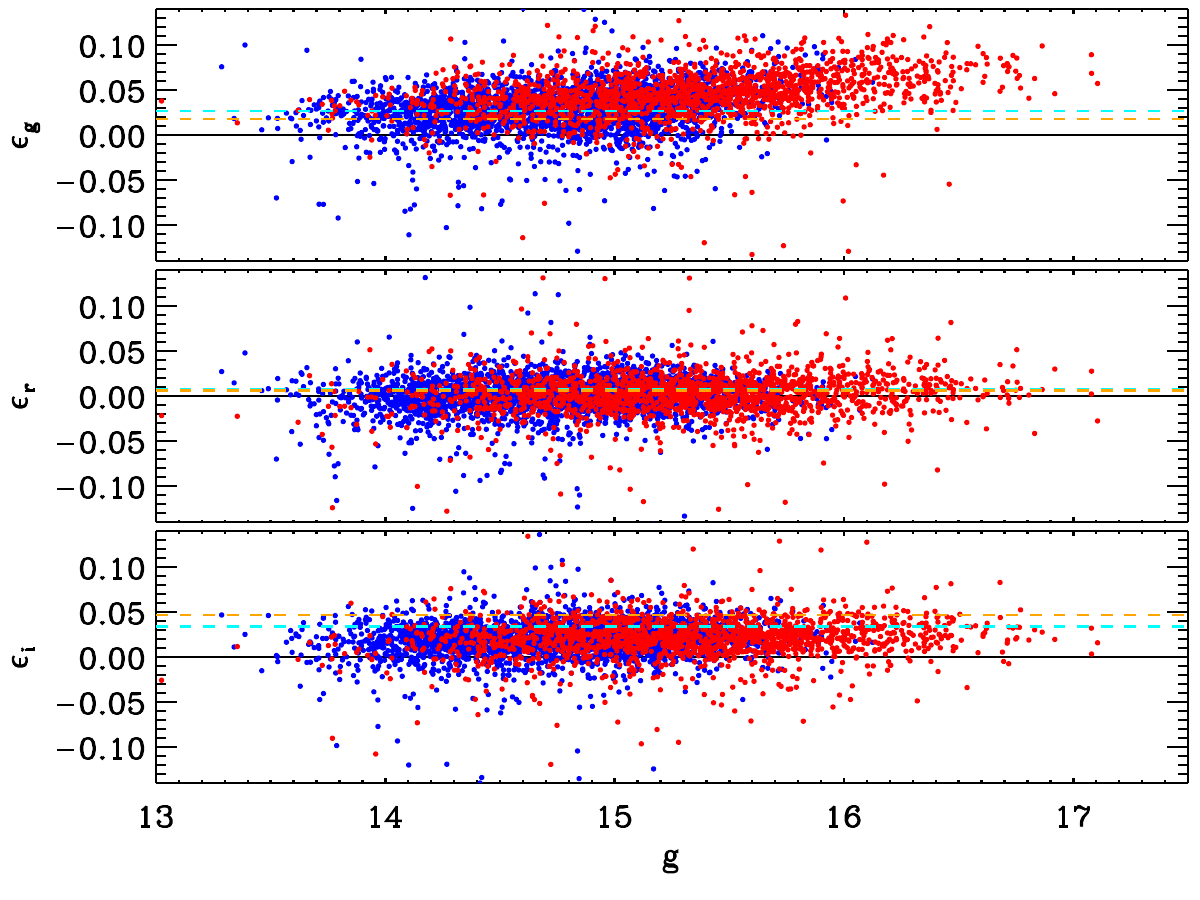}
    \caption{Same as Figure \ref{fig:xp_D10} but showing residuals as function of $g$ magnitudes.}
    \label{fig:xp2_D10}
\end{figure}

In the case of the $g$ band, the XP comparison reveals a clear colour-dependent offset that was not evident from the IRFM analysis alone, although the sign of the offset is consistent with what we found, and in disagreement with the CALSPEC comparison. Examining the residuals as a function of magnitude in Figure \ref{fig:xp2_D10} also reveals a noticeable magnitude-dependent trend in the $g$ band. This behaviour is not seen in the CALSPEC comparison but is in good agreement with the trends reported by \citet{Gaiadr3Syn2023A&A}. 

For the $i$ band, Gaia XP confirms the presence of an offset, albeit slightly smaller than the value inferred from the IRFM inversion.

\section{Conclusions}\label{sec:concl}

We have applied the Infrared Flux Method to derive photometric zero-points for the SDSS $ugriz$ system. Our analysis explores both the widely used transmission curves of \citet{Fukugita1996AJ} and the updated measurements of \citet{Doi2010AJ}. The stellar samples are drawn from the GALAH and APOGEE spectroscopic surveys, providing several thousand benchmark stars with accurate effective temperatures anchored to the IRFM scale.

Unlike most previous SDSS calibration studies, which typically rely on relatively few hot spectrophotometric standards such as white dwarfs, our analysis is based on thousands of late-type stars spanning a wide range of stellar parameters. This places our calibration on a stellar population that is directly relevant for Galactic archaeology studies, where the bulk of targets are FGK-type stars.

We find broadly consistent results when using the F96 and D10 transmission curves, with the exception of the $u$ band. When adopting the D10 response functions, which explicitly account for the red-leak, we derive significantly larger $u$-band zero-point offsets than those commonly quoted in the literature. Using the original F96 curves yields values closer to previous determinations, indicating that the discrepancy is driven by properly accounting for the leak. Our results therefore suggest that the $u$ band likely exhibits a colour-dependent zero-point offset.

For the remaining bands our results are largely consistent with previous studies. We confirm the excellent standardization of the $r$ band, with a zero-point very close to the AB system. The $i$ and $z$ bands show offsets of a few hundredths of a magnitude, broadly consistent with earlier estimates. In addition, we detect a small but statistically significant offset in the $g$ band, which has not been clearly identified in previous analyses.

Independent validation using CALSPEC spectrophotometric standards and Gaia XP spectra broadly supports the offsets derived from the IRFM inversion, particularly for the $r$ and $i$ bands. The comparison with Gaia XP spectra also reveals evidence for colour- and magnitude-dependent trends in the $g$ band, consistent with results reported for Gaia DR3 synthetic photometry.

The methodology presented here, which exploits the sensitivity of the IRFM to photometric zero-points using large stellar samples, provides a complementary approach to traditional spectrophotometric calibration techniques. This framework may prove particularly valuable for the calibration of future large-scale surveys such as LSST, CSST, and the Roman Space Telescope, where consistent cross-survey photometric standardization will be essential.

\section*{Acknowledgements}
We thank the referee, Dr. A.E. Lynas-Gary, for a careful and constructive report to strengthen the presentation. This work is supported by the Natural Science Foundation of China (NSFC) with grant Nos. 12125303, 12288102, 12090040/3. Part of this work has been carried out at the ANU Research School of Astronomy and Astrophysics, supported through the China Scholarship Council. This work is also supported by the National Key R\&D Program of China (grant No.2021YFA1600401/ 2021YFA1600403), the Yunnan Revitalization Talent Support Program—Science \& Technology Champion Project (No.202305AB350003), the New Cornerstone Science Foundation through the XPLORER PRIZE, and the International Centre of Supernovae, Yunnan Key Laboratory (No.202302AN360001). This work has made use of data from the European Space Agency (ESA) mission {\it Gaia} (\url{https://www.cosmos.esa.int/gaia}), processed by the {\it Gaia} Data Processing and Analysis Consortium (DPAC, \url{https://www.cosmos.esa.int/web/gaia/dpac/consortium}). Funding for the DPAC has been provided by national institutions, in particular the institutions participating in the {\it Gaia} Multilateral Agreement. This research has made use of NASA’s Astrophysics Data System, operated by the Smithsonian Astrophysical Observatory under NASA Cooperative Agreement 80NSSC21M0056. It also made use of TOPCAT, an interactive graphical viewer and editor for tabular data \citep{topcat2005ASPC}.

\section*{Data Availability Statements}
The data underlying this article are available in the article and in its online 
supplementary materials. The input stellar samples used in this study, including 
identifiers and relevant stellar parameters (effective temperature, surface 
gravity, and metallicity), as well as the photometric data required to reproduce the results, are provided in 
machine-readable CSV format. The original survey data are publicly available 
from the GALAH survey (\url{https://www.galah-survey.org}) and APOGEE 
(\url{https://www.sdss4.org/dr17/irspec/}).




\bibliographystyle{mnras}
\bibliography{ref} 

@ARTICLE{smith02,
       author = {{Smith}, J. Allyn and {Tucker}, Douglas L. and {Kent}, Stephen and {Richmond}, Michael W. and {Fukugita}, Masataka and {Ichikawa}, Takashi and {Ichikawa}, Shin-ichi and {Jorgensen}, Anders M. and {Uomoto}, Alan and {Gunn}, James E. and {Hamabe}, Masaru and {Watanabe}, Masaru and {Tolea}, Alin and {Henden}, Arne and {Annis}, James and {Pier}, Jeffrey R. and {McKay}, Timothy A. and {Brinkmann}, Jon and {Chen}, Bing and {Holtzman}, Jon and {Shimasaku}, Kazuhiro and {York}, Donald G.},
        title = "{The u'g'r'i'z' Standard-Star System}",
      journal = {\aj},
         year = 2002,
        month = apr,
       volume = {123},
       number = {4},
        pages = {2121-2144},
          doi = {10.1086/339311},
archivePrefix = {arXiv},
       eprint = {astro-ph/0201143},
 primaryClass = {astro-ph},
       adsurl = {https://ui.adsabs.harvard.edu/abs/2002AJ....123.2121S}
}

@ARTICLE{pancino,
       author = {{Pancino}, E. and {Sanna}, N. and {Altavilla}, G. and {Marinoni}, S. and {Rainer}, M. and {Cocozza}, G. and {Ragaini}, S. and {Galleti}, S. and {Bellazzini}, M. and {Bragaglia}, A. and {Tessicini}, G. and {Voss}, H. and {Carrasco}, J.~M. and {Jordi}, C. and {Harrison}, D.~L. and {De Angeli}, F. and {Evans}, D.~W. and {Fanari}, G.},
        title = "{The Gaia spectrophotometric standard stars survey - V. Preliminary flux tables for the calibration of Gaia DR2 and (E)DR3}",
      journal = {\mnras},
         year = 2021,
        month = may,
       volume = {503},
       number = {3},
        pages = {3660-3676},
          doi = {10.1093/mnras/stab766},
archivePrefix = {arXiv},
       eprint = {2103.07154},
 primaryClass = {astro-ph.SR},
       adsurl = {https://ui.adsabs.harvard.edu/abs/2021MNRAS.503.3660P}
}

@ARTICLE{c08,
       author = {{Casagrande}, Luca and {Flynn}, Chris and {Bessell}, Michael},
        title = "{M dwarfs: effective temperatures, radii and metallicities}",
      journal = {\mnras},
         year = 2008,
        month = sep,
       volume = {389},
       number = {2},
        pages = {585-607},
          doi = {10.1111/j.1365-2966.2008.13573.x},
archivePrefix = {arXiv},
       eprint = {0806.2471},
 primaryClass = {astro-ph},
       adsurl = {https://ui.adsabs.harvard.edu/abs/2008MNRAS.389..585C}
}

@ARTICLE{dahn02,
       author = {{Dahn}, Conard C. and {Harris}, Hugh C. and {Vrba}, Frederick J. and {Guetter}, Harry H. and {Canzian}, Blaise and {Henden}, Arne A. and {Levine}, Stephen E. and {Luginbuhl}, Christian B. and {Monet}, Alice K.~B. and {Monet}, David G. and {Pier}, Jeffrey R. and {Stone}, Ronald C. and {Walker}, Richard L. and {Burgasser}, Adam J. and {Gizis}, John E. and {Kirkpatrick}, J. Davy and {Liebert}, James and {Reid}, I. Neill},
        title = "{Astrometry and Photometry for Cool Dwarfs and Brown Dwarfs}",
      journal = {\aj},
         year = 2002,
        month = aug,
       volume = {124},
       number = {2},
        pages = {1170-1189},
          doi = {10.1086/341646},
archivePrefix = {arXiv},
       eprint = {astro-ph/0205050},
 primaryClass = {astro-ph},
       adsurl = {https://ui.adsabs.harvard.edu/abs/2002AJ....124.1170D}
}

@ARTICLE{sfd,
       author = {{Schlegel}, David J. and {Finkbeiner}, Douglas P. and {Davis}, Marc},
        title = "{Maps of Dust Infrared Emission for Use in Estimation of Reddening and Cosmic Microwave Background Radiation Foregrounds}",
      journal = {\apj},
         year = 1998,
        month = jun,
       volume = {500},
       number = {2},
        pages = {525-553},
          doi = {10.1086/305772},
archivePrefix = {arXiv},
       eprint = {astro-ph/9710327},
 primaryClass = {astro-ph},
       adsurl = {https://ui.adsabs.harvard.edu/abs/1998ApJ...500..525S}
}

@ARTICLE{SDSS_DR2,
       author = {{Abazajian}, Kevork and {Adelman-McCarthy}, Jennifer K. and {Ag{\"u}eros}, Marcel A. and {Allam}, Sahar S. and {Anderson}, Kurt and {Anderson}, Scott F. and {Annis}, James and {Bahcall}, Neta A. and {Baldry}, Ivan K. and {Bastian}, Steven and {Berlind}, Andreas and {Bernardi}, Mariangela and {Blanton}, Michael R. and {Bochanski}, Jr., John J. and {Boroski}, William N. and {Briggs}, John W. and {Brinkmann}, J. and {Brunner}, Robert J. and {Budav{\'a}ri}, Tam{\'a}s and {Carey}, Larry N. and {Carliles}, Samuel and {Castander}, Francisco J. and {Connolly}, A.~J. and {Csabai}, Istv{\'a}n and {Doi}, Mamoru and {Dong}, Feng and {Eisenstein}, Daniel J. and {Evans}, Michael L. and {Fan}, Xiaohui and {Finkbeiner}, Douglas P. and {Friedman}, Scott D. and {Frieman}, Joshua A. and {Fukugita}, Masataka and {Gal}, Roy R. and {Gillespie}, Bruce and {Glazebrook}, Karl and {Gray}, Jim and {Grebel}, Eva K. and {Gunn}, James E. and {Gurbani}, Vijay K. and {Hall}, Patrick B. and {Hamabe}, Masaru and {Harris}, Frederick H. and {Harris}, Hugh C. and {Harvanek}, Michael and {Heckman}, Timothy M. and {Hendry}, John S. and {Hennessy}, Gregory S. and {Hindsley}, Robert B. and {Hogan}, Craig J. and {Hogg}, David W. and {Holmgren}, Donald J. and {Ichikawa}, Shin-ichi and {Ichikawa}, Takashi and {Ivezi{\'c}}, {\v{Z}}eljko and {Jester}, Sebastian and {Johnston}, David E. and {Jorgensen}, Anders M. and {Kent}, Stephen M. and {Kleinman}, S.~J. and {Knapp}, G.~R. and {Kniazev}, Alexei Yu. and {Kron}, Richard G. and {Krzesinski}, Jurek and {Kunszt}, Peter Z. and {Kuropatkin}, Nickolai and {Lamb}, Donald Q. and {Lampeitl}, Hubert and {Lee}, Brian C. and {Leger}, R. French and {Li}, Nolan and {Lin}, Huan and {Loh}, Yeong-Shang and {Long}, Daniel C. and {Loveday}, Jon and {Lupton}, Robert H. and {Malik}, Tanu and {Margon}, Bruce and {Matsubara}, Takahiko and {McGehee}, Peregrine M. and {McKay}, Timothy A. and {Meiksin}, Avery and {Munn}, Jeffrey A. and {Nakajima}, Reiko and {Nash}, Thomas and {Neilsen}, Jr., Eric H. and {Newberg}, Heidi Jo and {Newman}, Peter R. and {Nichol}, Robert C. and {Nicinski}, Tom and {Nieto-Santisteban}, Maria and {Nitta}, Atsuko and {Okamura}, Sadanori and {O'Mullane}, William and {Ostriker}, Jeremiah P. and {Owen}, Russell and {Padmanabhan}, Nikhil and {Peoples}, John and {Pier}, Jeffrey R. and {Pope}, Adrian C. and {Quinn}, Thomas R. and {Richards}, Gordon T. and {Richmond}, Michael W. and {Rix}, Hans-Walter and {Rockosi}, Constance M. and {Schlegel}, David J. and {Schneider}, Donald P. and {Scranton}, Ryan and {Sekiguchi}, Maki and {Seljak}, Uros and {Sergey}, Gary and {Sesar}, Branimir and {Sheldon}, Erin and {Shimasaku}, Kazu and {Siegmund}, Walter A. and {Silvestri}, Nicole M. and {Smith}, J. Allyn and {Smol{\v{c}}i{\'c}}, Vernesa and {Snedden}, Stephanie A. and {Stebbins}, Albert and {Stoughton}, Chris and {Strauss}, Michael A. and {SubbaRao}, Mark and {Szalay}, Alexander S. and {Szapudi}, Istv{\'a}n and {Szkody}, Paula and {Szokoly}, Gyula P. and {Tegmark}, Max and {Teodoro}, Luis and {Thakar}, Aniruddha R. and {Tremonti}, Christy and {Tucker}, Douglas L. and {Uomoto}, Alan and {Vanden Berk}, Daniel E. and {Vandenberg}, Jan and {Vogeley}, Michael S. and {Voges}, Wolfgang and {Vogt}, Nicole P. and {Walkowicz}, Lucianne M. and {Wang}, Shu-i. and {Weinberg}, David H. and {West}, Andrew A. and {White}, Simon D.~M. and {Wilhite}, Brian C. and {Xu}, Yongzhong and {Yanny}, Brian and {Yasuda}, Naoki and {Yip}, Ching-Wa and {Yocum}, D.~R. and {York}, Donald G. and {Zehavi}, Idit and {Zibetti}, Stefano and {Zucker}, Daniel B.},
        title = "{The Second Data Release of the Sloan Digital Sky Survey}",
      journal = {\aj},
         year = 2004,
        month = jul,
       volume = {128},
       number = {1},
        pages = {502-512},
          doi = {10.1086/421365},
archivePrefix = {arXiv},
       eprint = {astro-ph/0403325},
 primaryClass = {astro-ph},
       adsurl = {https://ui.adsabs.harvard.edu/abs/2004AJ....128..502A}
}

@ARTICLE{Blackwell1977MNRAS,
       author = {{Blackwell}, D.~E. and {Shallis}, M.~J.},
        title = "{Stellar angular diameters from infrared photometry. Application to Arcturus and other stars; with effective temperatures.}",
      journal = {\mnras},
         year = 1977,
        month = jul,
       volume = {180},
        pages = {177-191},
          doi = {10.1093/mnras/180.2.177},
       adsurl = {https://ui.adsabs.harvard.edu/abs/1977MNRAS.180..177B}
}

@ARTICLE{Blackwell1979MNRAS,
       author = {{Blackwell}, D.~E. and {Shallis}, M.~J. and {Selby}, M.~J.},
        title = "{The infrared flux method for determining stellar angular diameters and effective temperatures.}",
      journal = {\mnras},
         year = 1979,
        month = sep,
       volume = {188},
        pages = {847-862},
          doi = {10.1093/mnras/188.4.847},
       adsurl = {https://ui.adsabs.harvard.edu/abs/1979MNRAS.188..847B}
}

@ARTICLE{Blackwell1980A&A,
       author = {{Blackwell}, D.~E. and {Petford}, A.~D. and {Shallis}, M.~J.},
        title = "{Use of the infra-red flux method for determining stellar effective temperatures and angular diameters; the stellar temperature scale.}",
      journal = {\aap},
         year = 1980,
        month = feb,
       volume = {82},
        pages = {249-252},
       adsurl = {https://ui.adsabs.harvard.edu/abs/1980A&A....82..249B}
}

@ARTICLE{Alonso1996A&AS,
       author = {{Alonso}, A. and {Arribas}, S. and {Martinez-Roger}, C.},
        title = "{Determination of effective temperatures for an extended sample of dwarfs and subdwarfs (F0-K5).}",
      journal = {\aaps},
         year = 1996,
        month = jun,
       volume = {117},
        pages = {227-254},
       adsurl = {https://ui.adsabs.harvard.edu/abs/1996A&AS..117..227A}
}

@ARTICLE{Casagrande2021MNRAS,
       author = {{Casagrande}, Luca and {Lin}, Jane and {Rains}, Adam D. and {Liu}, Fan and {Buder}, Sven and {Horner}, Jonathan and {Asplund}, Martin and {Lewis}, Geraint F. and {Martell}, Sarah L. and {Nordlander}, Thomas and {Stello}, Dennis and {Ting}, Yuan-Sen and {Wittenmyer}, Robert A. and {Bland-Hawthorn}, Joss and {Casey}, Andrew R. and {De Silva}, Gayandhi M. and {D'Orazi}, Valentina and {Freeman}, Ken C. and {Hayden}, Michael R. and {Kos}, Janez and {Lind}, Karin and {Schlesinger}, Katharine J. and {Sharma}, Sanjib and {Simpson}, Jeffrey D. and {Zucker}, Daniel B. and {Zwitter}, Toma{\v{z}}},
        title = "{The GALAH survey: effective temperature calibration from the InfraRed Flux Method in the Gaia system}",
      journal = {\mnras},
         year = 2021,
        month = oct,
       volume = {507},
       number = {2},
        pages = {2684-2696},
          doi = {10.1093/mnras/stab2304},
archivePrefix = {arXiv},
       eprint = {2011.02517},
 primaryClass = {astro-ph.SR},
       adsurl = {https://ui.adsabs.harvard.edu/abs/2021MNRAS.507.2684C}
}

@ARTICLE{Casagrande2010A&A,
       author = {{Casagrande}, L. and {Ram{\'\i}rez}, I. and {Mel{\'e}ndez}, J. and {Bessell}, M. and {Asplund}, M.},
        title = "{An absolutely calibrated T$_{eff}$ scale from the infrared flux method. Dwarfs and subgiants}",
      journal = {\aap},
         year = 2010,
        month = mar,
       volume = {512},
          eid = {A54},
        pages = {A54},
          doi = {10.1051/0004-6361/200913204},
archivePrefix = {arXiv},
       eprint = {1001.3142},
 primaryClass = {astro-ph.SR},
       adsurl = {https://ui.adsabs.harvard.edu/abs/2010A&A...512A..54C}
}

@ARTICLE{Fukugita1996AJ,
       author = {{Fukugita}, M. and {Ichikawa}, T. and {Gunn}, J.~E. and {Doi}, M. and {Shimasaku}, K. and {Schneider}, D.~P.},
        title = "{The Sloan Digital Sky Survey Photometric System}",
      journal = {\aj},
         year = 1996,
        month = apr,
       volume = {111},
        pages = {1748},
          doi = {10.1086/117915},
       adsurl = {https://ui.adsabs.harvard.edu/abs/1996AJ....111.1748F}
}

@ARTICLE{Doi2010AJ,
       author = {{Doi}, Mamoru and {Tanaka}, Masayuki and {Fukugita}, Masataka and {Gunn}, James E. and {Yasuda}, Naoki and {Ivezi{\'c}}, {\v{Z}}eljko and {Brinkmann}, Jon and {de Haars}, Ernst and {Kleinman}, S.~J. and {Krzesinski}, Jurek and {French Leger}, R.},
        title = "{Photometric Response Functions of the Sloan Digital Sky Survey Imager}",
      journal = {\aj},
         year = 2010,
        month = apr,
       volume = {139},
       number = {4},
        pages = {1628-1648},
          doi = {10.1088/0004-6256/139/4/1628},
archivePrefix = {arXiv},
       eprint = {1002.3701},
 primaryClass = {astro-ph.IM},
       adsurl = {https://ui.adsabs.harvard.edu/abs/2010AJ....139.1628D}
}

@ARTICLE{Casagrande2019MNRAS,
       author = {{Casagrande}, L. and {Wolf}, C. and {Mackey}, A.~D. and {Nordlander}, T. and {Yong}, D. and {Bessell}, M.},
        title = "{SkyMapper stellar parameters for Galactic Archaeology on a grand-scale}",
      journal = {\mnras},
         year = 2019,
        month = jan,
       volume = {482},
       number = {2},
        pages = {2770-2787},
          doi = {10.1093/mnras/sty2878},
archivePrefix = {arXiv},
       eprint = {1810.09581},
 primaryClass = {astro-ph.SR},
       adsurl = {https://ui.adsabs.harvard.edu/abs/2019MNRAS.482.2770C}
}

@ARTICLE{York2000AJ,
       author = {{York}, Donald G. and {Adelman}, J. and {Anderson}, John E., Jr. and {Anderson}, Scott F. and {Annis}, James and {Bahcall}, Neta A. and {Bakken}, J.~A. and {Barkhouser}, Robert and {Bastian}, Steven and {Berman}, Eileen and {Boroski}, William N. and {Bracker}, Steve and {Briegel}, Charlie and {Briggs}, John W. and {Brinkmann}, J. and {Brunner}, Robert and {Burles}, Scott and {Carey}, Larry and {Carr}, Michael A. and {Castander}, Francisco J. and {Chen}, Bing and {Colestock}, Patrick L. and {Connolly}, A.~J. and {Crocker}, J.~H. and {Csabai}, Istv{\'a}n and {Czarapata}, Paul C. and {Davis}, John Eric and {Doi}, Mamoru and {Dombeck}, Tom and {Eisenstein}, Daniel and {Ellman}, Nancy and {Elms}, Brian R. and {Evans}, Michael L. and {Fan}, Xiaohui and {Federwitz}, Glenn R. and {Fiscelli}, Larry and {Friedman}, Scott and {Frieman}, Joshua A. and {Fukugita}, Masataka and {Gillespie}, Bruce and {Gunn}, James E. and {Gurbani}, Vijay K. and {de Haas}, Ernst and {Haldeman}, Merle and {Harris}, Frederick H. and {Hayes}, J. and {Heckman}, Timothy M. and {Hennessy}, G.~S. and {Hindsley}, Robert B. and {Holm}, Scott and {Holmgren}, Donald J. and {Huang}, Chi-hao and {Hull}, Charles and {Husby}, Don and {Ichikawa}, Shin-Ichi and {Ichikawa}, Takashi and {Ivezi{\'c}}, {\v{Z}}eljko and {Kent}, Stephen and {Kim}, Rita S.~J. and {Kinney}, E. and {Klaene}, Mark and {Kleinman}, A.~N. and {Kleinman}, S. and {Knapp}, G.~R. and {Korienek}, John and {Kron}, Richard G. and {Kunszt}, Peter Z. and {Lamb}, D.~Q. and {Lee}, B. and {Leger}, R. French and {Limmongkol}, Siriluk and {Lindenmeyer}, Carl and {Long}, Daniel C. and {Loomis}, Craig and {Loveday}, Jon and {Lucinio}, Rich and {Lupton}, Robert H. and {MacKinnon}, Bryan and {Mannery}, Edward J. and {Mantsch}, P.~M. and {Margon}, Bruce and {McGehee}, Peregrine and {McKay}, Timothy A. and {Meiksin}, Avery and {Merelli}, Aronne and {Monet}, David G. and {Munn}, Jeffrey A. and {Narayanan}, Vijay K. and {Nash}, Thomas and {Neilsen}, Eric and {Neswold}, Rich and {Newberg}, Heidi Jo and {Nichol}, R.~C. and {Nicinski}, Tom and {Nonino}, Mario and {Okada}, Norio and {Okamura}, Sadanori and {Ostriker}, Jeremiah P. and {Owen}, Russell and {Pauls}, A. George and {Peoples}, John and {Peterson}, R.~L. and {Petravick}, Donald and {Pier}, Jeffrey R. and {Pope}, Adrian and {Pordes}, Ruth and {Prosapio}, Angela and {Rechenmacher}, Ron and {Quinn}, Thomas R. and {Richards}, Gordon T. and {Richmond}, Michael W. and {Rivetta}, Claudio H. and {Rockosi}, Constance M. and {Ruthmansdorfer}, Kurt and {Sandford}, Dale and {Schlegel}, David J. and {Schneider}, Donald P. and {Sekiguchi}, Maki and {Sergey}, Gary and {Shimasaku}, Kazuhiro and {Siegmund}, Walter A. and {Smee}, Stephen and {Smith}, J. Allyn and {Snedden}, S. and {Stone}, R. and {Stoughton}, Chris and {Strauss}, Michael A. and {Stubbs}, Christopher and {SubbaRao}, Mark and {Szalay}, Alexander S. and {Szapudi}, Istvan and {Szokoly}, Gyula P. and {Thakar}, Anirudda R. and {Tremonti}, Christy and {Tucker}, Douglas L. and {Uomoto}, Alan and {Vanden Berk}, Dan and {Vogeley}, Michael S. and {Waddell}, Patrick and {Wang}, Shu-i. and {Watanabe}, Masaru and {Weinberg}, David H. and {Yanny}, Brian and {Yasuda}, Naoki and {SDSS Collaboration}},
        title = "{The Sloan Digital Sky Survey: Technical Summary}",
      journal = {\aj},
         year = 2000,
        month = sep,
       volume = {120},
       number = {3},
        pages = {1579-1587},
          doi = {10.1086/301513},
archivePrefix = {arXiv},
       eprint = {astro-ph/0006396},
 primaryClass = {astro-ph},
       adsurl = {https://ui.adsabs.harvard.edu/abs/2000AJ....120.1579Y}
}

@ARTICLE{Lupton1999AJ,
       author = {{Lupton}, Robert H. and {Gunn}, James E. and {Szalay}, Alexander S.},
        title = "{A Modified Magnitude System that Produces Well-Behaved Magnitudes, Colors, and Errors Even for Low Signal-to-Noise Ratio Measurements}",
      journal = {\aj},
         year = 1999,
        month = sep,
       volume = {118},
       number = {3},
        pages = {1406-1410},
          doi = {10.1086/301004},
archivePrefix = {arXiv},
       eprint = {astro-ph/9903081},
 primaryClass = {astro-ph},
       adsurl = {https://ui.adsabs.harvard.edu/abs/1999AJ....118.1406L}
}

@INPROCEEDINGS{topcat2005ASPC,
       author = {{Taylor}, M.~B.},
        title = "{TOPCAT \& STIL: Starlink Table/VOTable Processing Software}",
    booktitle = {Astronomical Data Analysis Software and Systems XIV},
         year = 2005,
       editor = {{Shopbell}, P. and {Britton}, M. and {Ebert}, R.},
       series = {Astronomical Society of the Pacific Conference Series},
       volume = {347},
        month = dec,
        pages = {29},
       adsurl = {https://ui.adsabs.harvard.edu/abs/2005ASPC..347...29T}
}

@ARTICLE{BBSuzuki2018AJ,
       author = {{Suzuki}, Nao and {Fukugita}, Masataka},
        title = "{Blackbody Stars}",
      journal = {\aj},
         year = 2018,
        month = nov,
       volume = {156},
       number = {5},
          eid = {219},
        pages = {219},
          doi = {10.3847/1538-3881/aac88b},
archivePrefix = {arXiv},
       eprint = {1711.01122},
 primaryClass = {astro-ph.SR},
       adsurl = {https://ui.adsabs.harvard.edu/abs/2018AJ....156..219S}
}

@ARTICLE{Oke1983ApJ,
       author = {{Oke}, J.~B. and {Gunn}, J.~E.},
        title = "{Secondary standard stars for absolute spectrophotometry.}",
      journal = {\apj},
         year = 1983,
        month = mar,
       volume = {266},
        pages = {713-717},
          doi = {10.1086/160817},
       adsurl = {https://ui.adsabs.harvard.edu/abs/1983ApJ...266..713O}
}

@ARTICLE{Casagrande2014MNRAS,
       author = {{Casagrande}, L. and {VandenBerg}, Don A.},
        title = "{Synthetic stellar photometry - I. General considerations and new transformations for broad-band systems}",
      journal = {\mnras},
         year = 2014,
        month = oct,
       volume = {444},
       number = {1},
        pages = {392-419},
          doi = {10.1093/mnras/stu1476},
archivePrefix = {arXiv},
       eprint = {1407.6095},
 primaryClass = {astro-ph.SR},
       adsurl = {https://ui.adsabs.harvard.edu/abs/2014MNRAS.444..392C}
}

@ARTICLE{Eisenstein2006ApJS,
       author = {{Eisenstein}, Daniel J. and {Liebert}, James and {Harris}, Hugh C. and {Kleinman}, S.~J. and {Nitta}, Atsuko and {Silvestri}, Nicole and {Anderson}, Scott A. and {Barentine}, J.~C. and {Brewington}, Howard J. and {Brinkmann}, J. and {Harvanek}, Michael and {Krzesi{\'n}ski}, Jurek and {Neilsen}, Eric H., Jr. and {Long}, Dan and {Schneider}, Donald P. and {Snedden}, Stephanie A.},
        title = "{A Catalog of Spectroscopically Confirmed White Dwarfs from the Sloan Digital Sky Survey Data Release 4}",
      journal = {\apjs},
         year = 2006,
        month = nov,
       volume = {167},
       number = {1},
        pages = {40-58},
          doi = {10.1086/507110},
archivePrefix = {arXiv},
       eprint = {astro-ph/0606700},
 primaryClass = {astro-ph},
       adsurl = {https://ui.adsabs.harvard.edu/abs/2006ApJS..167...40E}
}

@ARTICLE{Blackwell1991A&A,
       author = {{Blackwell}, D.~E. and {Lynas-Gray}, A.~E. and {Petford}, A.~D.},
        title = "{Effect of improved H- opacity on the infrared flux method temperature scale and derived angular diameters. Use of a self-consistent calibration.}",
      journal = {\aap},
         year = 1991,
        month = may,
       volume = {245},
        pages = {567},
       adsurl = {https://ui.adsabs.harvard.edu/abs/1991A&A...245..567B}
}

@ARTICLE{Casagrande2006MNRAS,
       author = {{Casagrande}, Luca and {Portinari}, Laura and {Flynn}, Chris},
        title = "{Accurate fundamental parameters for lower main-sequence stars}",
      journal = {\mnras},
         year = 2006,
        month = nov,
       volume = {373},
       number = {1},
        pages = {13-44},
          doi = {10.1111/j.1365-2966.2006.10999.x},
archivePrefix = {arXiv},
       eprint = {astro-ph/0608504},
 primaryClass = {astro-ph},
       adsurl = {https://ui.adsabs.harvard.edu/abs/2006MNRAS.373...13C}
}

@ARTICLE{Bohlin2014AJ,
       author = {{Bohlin}, R.~C.},
        title = "{Hubble Space Telescope CALSPEC Flux Standards: Sirius (and Vega)}",
      journal = {\aj},
         year = 2014,
        month = jun,
       volume = {147},
       number = {6},
          eid = {127},
        pages = {127},
          doi = {10.1088/0004-6256/147/6/127},
       adsurl = {https://ui.adsabs.harvard.edu/abs/2014AJ....147..127B}
}

@ARTICLE{Bohlin2001AJ,
       author = {{Bohlin}, R.~C. and {Dickinson}, M.~E. and {Calzetti}, D.},
        title = "{Spectrophotometric Standards from the Far-Ultraviolet to the Near-Infrared: STIS and NICMOS Fluxes}",
      journal = {\aj},
         year = 2001,
        month = oct,
       volume = {122},
       number = {4},
        pages = {2118-2128},
          doi = {10.1086/323137},
       adsurl = {https://ui.adsabs.harvard.edu/abs/2001AJ....122.2118B}
}

@MISC{Bohlin2022stis,
       author = {{Bohlin}, Ralph C. and {Lockwood}, Sean},
        title = "{Update of the STIS CTE Correction Formula for Stellar Spectra}",
 howpublished = {Instrument Science Report STIS 2022-7, 11 pages},
         year = 2022,
        month = oct,
        pages = {7},
       adsurl = {https://ui.adsabs.harvard.edu/abs/2022stis.rept....7B}
}

@ARTICLE{Gaiadr3Syn2023A&A,
       author = {{Gaia Collaboration} and {Montegriffo}, P. and {Bellazzini}, M. and {De Angeli}, F. and {Andrae}, R. and {Barstow}, M.~A. and {Bossini}, D. and {Bragaglia}, A. and {Burgess}, P.~W. and {Cacciari}, C. and {Carrasco}, J.~M. and {Chornay}, N. and {Delchambre}, L. and {Evans}, D.~W. and {Fouesneau}, M. and {Fr{\'e}mat}, Y. and {Garabato}, D. and {Jordi}, C. and {Manteiga}, M. and {Massari}, D. and {Palaversa}, L. and {Pancino}, E. and {Riello}, M. and {Ruz Mieres}, D. and {Sanna}, N. and {Santove{\~n}a}, R. and {Sordo}, R. and {Vallenari}, A. and {Walton}, N.~A. and {Brown}, A.~G.~A. and {Prusti}, T. and {de Bruijne}, J.~H.~J. and {Arenou}, F. and {Babusiaux}, C. and {Biermann}, M. and {Creevey}, O.~L. and {Ducourant}, C. and {Eyer}, L. and {Guerra}, R. and {Hutton}, A. and {Klioner}, S.~A. and {Lammers}, U.~L. and {Lindegren}, L. and {Luri}, X. and {Mignard}, F. and {Panem}, C. and {Pourbaix}, D. and {Randich}, S. and {Sartoretti}, P. and {Soubiran}, C. and {Tanga}, P. and {Bailer-Jones}, C.~A.~L. and {Bastian}, U. and {Drimmel}, R. and {Jansen}, F. and {Katz}, D. and {Lattanzi}, M.~G. and {van Leeuwen}, F. and {Bakker}, J. and {Casta{\~n}eda}, J. and {Fabricius}, C. and {Galluccio}, L. and {Guerrier}, A. and {Heiter}, U. and {Masana}, E. and {Messineo}, R. and {Mowlavi}, N. and {Nicolas}, C. and {Nienartowicz}, K. and {Pailler}, F. and {Panuzzo}, P. and {Riclet}, F. and {Roux}, W. and {Seabroke}, G.~M. and {Th{\'e}venin}, F. and {Gracia-Abril}, G. and {Portell}, J. and {Teyssier}, D. and {Altmann}, M. and {Audard}, M. and {Bellas-Velidis}, I. and {Benson}, K. and {Berthier}, J. and {Blomme}, R. and {Busonero}, D. and {Busso}, G. and {C{\'a}novas}, H. and {Carry}, B. and {Cellino}, A. and {Cheek}, N. and {Clementini}, G. and {Damerdji}, Y. and {Davidson}, M. and {de Teodoro}, P. and {Nu{\~n}ez Campos}, M. and {Dell'Oro}, A. and {Esquej}, P. and {Fern{\'a}ndez-Hern{\'a}ndez}, J. and {Fraile}, E. and {Garc{\'\i}a-Lario}, P. and {Gosset}, E. and {Haigron}, R. and {Halbwachs}, J. -L. and {Hambly}, N.~C. and {Harrison}, D.~L. and {Hern{\'a}ndez}, J. and {Hestroffer}, D. and {Hodgkin}, S.~T. and {Holl}, B. and {Jan{\ss}en}, K. and {Jevardat de Fombelle}, G. and {Jordan}, S. and {Krone-Martins}, A. and {Lanzafame}, A.~C. and {L{\"o}ffler}, W. and {Marchal}, O. and {Marrese}, P.~M. and {Moitinho}, A. and {Muinonen}, K. and {Osborne}, P. and {Pauwels}, T. and {Recio-Blanco}, A. and {Reyl{\'e}}, C. and {Rimoldini}, L. and {Roegiers}, T. and {Rybizki}, J. and {Sarro}, L.~M. and {Siopis}, C. and {Smith}, M. and {Sozzetti}, A. and {Utrilla}, E. and {van Leeuwen}, M. and {Abbas}, U. and {{\'A}brah{\'a}m}, P. and {Abreu Aramburu}, A. and {Aerts}, C. and {Aguado}, J.~J. and {Ajaj}, M. and {Aldea-Montero}, F. and {Altavilla}, G. and {{\'A}lvarez}, M.~A. and {Alves}, J. and {Anderson}, R.~I. and {Anglada Varela}, E. and {Antoja}, T. and {Baines}, D. and {Baker}, S.~G. and {Balaguer-N{\'u}{\~n}ez}, L. and {Balbinot}, E. and {Balog}, Z. and {Barache}, C. and {Barbato}, D. and {Barros}, M. and {Bartolom{\'e}}, S. and {Bassilana}, J. -L. and {Bauchet}, N. and {Becciani}, U. and {Berihuete}, A. and {Bernet}, M. and {Bertone}, S. and {Bianchi}, L. and {Binnenfeld}, A. and {Blanco-Cuaresma}, S. and {Boch}, T. and {Bombrun}, A. and {Bouquillon}, S. and {Bramante}, L. and {Breedt}, E. and {Bressan}, A. and {Brouillet}, N. and {Brugaletta}, E. and {Bucciarelli}, B. and {Burlacu}, A. and {Butkevich}, A.~G. and {Buzzi}, R. and {Caffau}, E. and {Cancelliere}, R. and {Cantat-Gaudin}, T. and {Carballo}, R. and {Carlucci}, T. and {Carnerero}, M.~I. and {Casamiquela}, L. and {Castellani}, M. and {Castro-Ginard}, A. and {Chaoul}, L. and {Charlot}, P. and {Chemin}, L. and {Chiaramida}, V. and {Chiavassa}, A. and {Comoretto}, G. and {Contursi}, G. and {Cooper}, W.~J. and {Cornez}, T. and {Cowell}, S. and {Crifo}, F. and {Cropper}, M. and {Crosta}, M. and {Crowley}, C. and {Dafonte}, C. and {Dapergolas}, A. and {David}, P. and {de Laverny}, P. and {De Luise}, F. and {De March}, R. and {De Ridder}, J. and {de Souza}, R. and {de Torres}, A. and {del Peloso}, E.~F. and {del Pozo}, E. and {Delbo}, M. and {Delgado}, A. and {Delisle}, J. -B. and {Demouchy}, C. and {Dharmawardena}, T.~E. and {Diakite}, S. and {Diener}, C. and {Distefano}, E. and {Dolding}, C. and {Enke}, H. and {Fabre}, C. and {Fabrizio}, M. and {Faigler}, S. and {Fedorets}, G. and {Fernique}, P. and {Figueras}, F. and {Fournier}, Y. and {Fouron}, C. and {Fragkoudi}, F. and {Gai}, M. and {Garcia-Gutierrez}, A. and {Garcia-Reinaldos}, M. and {Garc{\'\i}a-Torres}, M. and {Garofalo}, A. and {Gavel}, A. and {Gavras}, P. and {Gerlach}, E. and {Geyer}, R. and {Giacobbe}, P. and {Gilmore}, G. and {Girona}, S. and {Giuffrida}, G. and {Gomel}, R. and {Gomez}, A. and {Gonz{\'a}lez-N{\'u}{\~n}ez}, J. and {Gonz{\'a}lez-Santamar{\'\i}a}, I. and {Gonz{\'a}lez-Vidal}, J.~J. and {Granvik}, M. and {Guillout}, P. and {Guiraud}, J. and {Guti{\'e}rrez-S{\'a}nchez}, R. and {Guy}, L.~P. and {Hatzidimitriou}, D. and {Hauser}, M. and {Haywood}, M. and {Helmer}, A. and {Helmi}, A. and {Sarmiento}, M.~H. and {Hidalgo}, S.~L. and {H{\l}adczuk}, N. and {Hobbs}, D. and {Holland}, G. and {Huckle}, H.~E. and {Jardine}, K. and {Jasniewicz}, G. and {Jean-Antoine Piccolo}, A. and {Jim{\'e}nez-Arranz}, {\'O}. and {Juaristi Campillo}, J. and {Julbe}, F. and {Karbevska}, L. and {Kervella}, P. and {Khanna}, S. and {Kordopatis}, G. and {Korn}, A.~J. and {K{\'o}sp{\'a}l}, {\'A}. and {Kostrzewa-Rutkowska}, Z. and {Kruszy{\'n}ska}, K. and {Kun}, M. and {Laizeau}, P. and {Lambert}, S. and {Lanza}, A.~F. and {Lasne}, Y. and {Le Campion}, J. -F. and {Lebreton}, Y. and {Lebzelter}, T. and {Leccia}, S. and {Leclerc}, N. and {Lecoeur-Taibi}, I. and {Liao}, S. and {Licata}, E.~L. and {Lindstr{\'o}m}, H.~E.~P. and {Lister}, T.~A. and {Livanou}, E. and {Lobel}, A. and {Lorca}, A. and {Loup}, C. and {Madrero Pardo}, P. and {Magdaleno Romeo}, A. and {Managau}, S. and {Mann}, R.~G. and {Marchant}, J.~M. and {Marconi}, M. and {Marcos}, J. and {Marcos Santos}, M.~M.~S. and {Mar{\'\i}n Pina}, D. and {Marinoni}, S. and {Marocco}, F. and {Marshall}, D.~J. and {Martin Polo}, L. and {Mart{\'\i}n-Fleitas}, J.~M. and {Marton}, G. and {Mary}, N. and {Masip}, A. and {Mastrobuono-Battisti}, A. and {Mazeh}, T. and {McMillan}, P.~J. and {Messina}, S. and {Michalik}, D. and {Millar}, N.~R. and {Mints}, A. and {Molina}, D. and {Molinaro}, R. and {Moln{\'a}r}, L. and {Monari}, G. and {Mongui{\'o}}, M. and {Montero}, A. and {Mor}, R. and {Mora}, A. and {Morbidelli}, R. and {Morel}, T. and {Morris}, D. and {Muraveva}, T. and {Murphy}, C.~P. and {Musella}, I. and {Nagy}, Z. and {Noval}, L. and {Oca{\~n}a}, F. and {Ogden}, A. and {Ordenovic}, C. and {Osinde}, J.~O. and {Pagani}, C. and {Pagano}, I. and {Palicio}, P.~A. and {Pallas-Quintela}, L. and {Panahi}, A. and {Payne-Wardenaar}, S. and {Pe{\~n}alosa Esteller}, X. and {Penttil{\"a}}, A. and {Pichon}, B. and {Piersimoni}, A.~M. and {Pineau}, F. -X. and {Plachy}, E. and {Plum}, G. and {Poggio}, E. and {Pr{\v{s}}a}, A. and {Pulone}, L. and {Racero}, E. and {Ragaini}, S. and {Rainer}, M. and {Raiteri}, C.~M. and {Ramos}, P. and {Ramos-Lerate}, M. and {Re Fiorentin}, P. and {Regibo}, S. and {Richards}, P.~J. and {Rios Diaz}, C. and {Ripepi}, V. and {Riva}, A. and {Rix}, H. -W. and {Rixon}, G. and {Robichon}, N. and {Robin}, A.~C. and {Robin}, C. and {Roelens}, M. and {Rogues}, H.~R.~O. and {Rohrbasser}, L. and {Romero-G{\'o}mez}, M. and {Rowell}, N. and {Royer}, F. and {Rybicki}, K.~A. and {Sadowski}, G. and {S{\'a}ez N{\'u}{\~n}ez}, A. and {Sagrist{\`a} Sell{\'e}s}, A. and {Sahlmann}, J. and {Salguero}, E. and {Samaras}, N. and {Sanchez Gimenez}, V. and {Sarasso}, M. and {Schultheis}, M.~S. and {Sciacca}, E. and {Segol}, M. and {Segovia}, J.~C. and {S{\'e}gransan}, D. and {Semeux}, D. and {Shahaf}, S. and {Siddiqui}, H.~I. and {Siebert}, A. and {Siltala}, L. and {Silvelo}, A. and {Slezak}, E. and {Slezak}, I. and {Smart}, R.~L. and {Snaith}, O.~N. and {Solano}, E. and {Solitro}, F. and {Souami}, D. and {Souchay}, J. and {Spagna}, A. and {Spina}, L. and {Spoto}, F. and {Steele}, I.~A. and {Steidelm{\"u}ller}, H. and {Stephenson}, C.~A. and {S{\"u}veges}, M. and {Surdej}, J. and {Szabados}, L. and {Szegedi-Elek}, E. and {Taris}, F. and {Taylor}, M.~B. and {Teixeira}, R. and {Tolomei}, L. and {Tonello}, N. and {Torra}, F. and {Torra}, J. and {Torralba Elipe}, G. and {Trabucchi}, M. and {Tsounis}, A.~T. and {Turon}, C. and {Ulla}, A. and {Unger}, N. and {Vaillant}, M.~V. and {van Dillen}, E. and {van Reeven}, W. and {Vanel}, O. and {Vecchiato}, A. and {Viala}, Y. and {Vicente}, D. and {Voutsinas}, S. and {Wevers}, T. and {Wyrzykowski}, {\L}. and {Yoldas}, A. and {Yvard}, P. and {Zhao}, H. and {Zorec}, J. and {Zucker}, S. and {Zwitter}, T.},
        title = "{Gaia Data Release 3. The Galaxy in your preferred colours: Synthetic photometry from Gaia low-resolution spectra}",
      journal = {\aap},
         year = 2023,
        month = jun,
       volume = {674},
          eid = {A33},
        pages = {A33},
          doi = {10.1051/0004-6361/202243709},
archivePrefix = {arXiv},
       eprint = {2206.06215},
 primaryClass = {astro-ph.SR},
       adsurl = {https://ui.adsabs.harvard.edu/abs/2023A&A...674A..33G}
}

@ARTICLE{Holberg2006AJ,
       author = {{Holberg}, J.~B. and {Bergeron}, Pierre},
        title = "{Calibration of Synthetic Photometry Using DA White Dwarfs}",
      journal = {\aj},
         year = 2006,
        month = sep,
       volume = {132},
       number = {3},
        pages = {1221-1233},
          doi = {10.1086/505938},
       adsurl = {https://ui.adsabs.harvard.edu/abs/2006AJ....132.1221H}
}

@ARTICLE{Blackwell1994A&A,
       author = {{Blackwell}, D.~E. and {Lynas-Gray}, A.~E.},
        title = "{Stellar effective temperatures and angular diameters determined by the infrared flux method (IRFM) : revisions using improved Kurucz LTE stellar atmospheres.}",
      journal = {\aap},
         year = 1994,
        month = feb,
       volume = {282},
        pages = {899-910},
       adsurl = {https://ui.adsabs.harvard.edu/abs/1994A&A...282..899B}
}

@ARTICLE{Bohlin2025AJ,
       author = {{Bohlin}, Ralph C. and {Deustua}, Susana and {Narayan}, Gautham and {Saha}, Abhijit and {Calamida}, Annalisa and {Gordon}, Karl D. and {Holberg}, Jay B. and {Hubeny}, Ivan and {Matheson}, Thomas and {Rest}, Armin},
        title = "{Faint White Dwarf Flux Standards: Data and Models}",
      journal = {\aj},
         year = 2025,
        month = jan,
       volume = {169},
       number = {1},
          eid = {40},
        pages = {40},
          doi = {10.3847/1538-3881/ad93d8},
archivePrefix = {arXiv},
       eprint = {2411.09049},
 primaryClass = {astro-ph.IM},
       adsurl = {https://ui.adsabs.harvard.edu/abs/2025AJ....169...40B}
}

@ARTICLE{LSST2019ApJ,
       author = {{Ivezi{\'c}}, {\v{Z}}eljko and {Kahn}, Steven M. and {Tyson}, J. Anthony and {Abel}, Bob and {Acosta}, Emily and {Allsman}, Robyn and {Alonso}, David and {AlSayyad}, Yusra and {Anderson}, Scott F. and {Andrew}, John and {Angel}, James Roger P. and {Angeli}, George Z. and {Ansari}, Reza and {Antilogus}, Pierre and {Araujo}, Constanza and {Armstrong}, Robert and {Arndt}, Kirk T. and {Astier}, Pierre and {Aubourg}, {\'E}ric and {Auza}, Nicole and {Axelrod}, Tim S. and {Bard}, Deborah J. and {Barr}, Jeff D. and {Barrau}, Aurelian and {Bartlett}, James G. and {Bauer}, Amanda E. and {Bauman}, Brian J. and {Baumont}, Sylvain and {Bechtol}, Ellen and {Bechtol}, Keith and {Becker}, Andrew C. and {Becla}, Jacek and {Beldica}, Cristina and {Bellavia}, Steve and {Bianco}, Federica B. and {Biswas}, Rahul and {Blanc}, Guillaume and {Blazek}, Jonathan and {Blandford}, Roger D. and {Bloom}, Josh S. and {Bogart}, Joanne and {Bond}, Tim W. and {Booth}, Michael T. and {Borgland}, Anders W. and {Borne}, Kirk and {Bosch}, James F. and {Boutigny}, Dominique and {Brackett}, Craig A. and {Bradshaw}, Andrew and {Brandt}, William Nielsen and {Brown}, Michael E. and {Bullock}, James S. and {Burchat}, Patricia and {Burke}, David L. and {Cagnoli}, Gianpietro and {Calabrese}, Daniel and {Callahan}, Shawn and {Callen}, Alice L. and {Carlin}, Jeffrey L. and {Carlson}, Erin L. and {Chandrasekharan}, Srinivasan and {Charles-Emerson}, Glenaver and {Chesley}, Steve and {Cheu}, Elliott C. and {Chiang}, Hsin-Fang and {Chiang}, James and {Chirino}, Carol and {Chow}, Derek and {Ciardi}, David R. and {Claver}, Charles F. and {Cohen-Tanugi}, Johann and {Cockrum}, Joseph J. and {Coles}, Rebecca and {Connolly}, Andrew J. and {Cook}, Kem H. and {Cooray}, Asantha and {Covey}, Kevin R. and {Cribbs}, Chris and {Cui}, Wei and {Cutri}, Roc and {Daly}, Philip N. and {Daniel}, Scott F. and {Daruich}, Felipe and {Daubard}, Guillaume and {Daues}, Greg and {Dawson}, William and {Delgado}, Francisco and {Dellapenna}, Alfred and {de Peyster}, Robert and {de Val-Borro}, Miguel and {Digel}, Seth W. and {Doherty}, Peter and {Dubois}, Richard and {Dubois-Felsmann}, Gregory P. and {Durech}, Josef and {Economou}, Frossie and {Eifler}, Tim and {Eracleous}, Michael and {Emmons}, Benjamin L. and {Fausti Neto}, Angelo and {Ferguson}, Henry and {Figueroa}, Enrique and {Fisher-Levine}, Merlin and {Focke}, Warren and {Foss}, Michael D. and {Frank}, James and {Freemon}, Michael D. and {Gangler}, Emmanuel and {Gawiser}, Eric and {Geary}, John C. and {Gee}, Perry and {Geha}, Marla and {Gessner}, Charles J.~B. and {Gibson}, Robert R. and {Gilmore}, D. Kirk and {Glanzman}, Thomas and {Glick}, William and {Goldina}, Tatiana and {Goldstein}, Daniel A. and {Goodenow}, Iain and {Graham}, Melissa L. and {Gressler}, William J. and {Gris}, Philippe and {Guy}, Leanne P. and {Guyonnet}, Augustin and {Haller}, Gunther and {Harris}, Ron and {Hascall}, Patrick A. and {Haupt}, Justine and {Hernandez}, Fabio and {Herrmann}, Sven and {Hileman}, Edward and {Hoblitt}, Joshua and {Hodgson}, John A. and {Hogan}, Craig and {Howard}, James D. and {Huang}, Dajun and {Huffer}, Michael E. and {Ingraham}, Patrick and {Innes}, Walter R. and {Jacoby}, Suzanne H. and {Jain}, Bhuvnesh and {Jammes}, Fabrice and {Jee}, M. James and {Jenness}, Tim and {Jernigan}, Garrett and {Jevremovi{\'c}}, Darko and {Johns}, Kenneth and {Johnson}, Anthony S. and {Johnson}, Margaret W.~G. and {Jones}, R. Lynne and {Juramy-Gilles}, Claire and {Juri{\'c}}, Mario and {Kalirai}, Jason S. and {Kallivayalil}, Nitya J. and {Kalmbach}, Bryce and {Kantor}, Jeffrey P. and {Karst}, Pierre and {Kasliwal}, Mansi M. and {Kelly}, Heather and {Kessler}, Richard and {Kinnison}, Veronica and {Kirkby}, David and {Knox}, Lloyd and {Kotov}, Ivan V. and {Krabbendam}, Victor L. and {Krughoff}, K. Simon and {Kub{\'a}nek}, Petr and {Kuczewski}, John and {Kulkarni}, Shri and {Ku}, John and {Kurita}, Nadine R. and {Lage}, Craig S. and {Lambert}, Ron and {Lange}, Travis and {Langton}, J. Brian and {Le Guillou}, Laurent and {Levine}, Deborah and {Liang}, Ming and {Lim}, Kian-Tat and {Lintott}, Chris J. and {Long}, Kevin E. and {Lopez}, Margaux and {Lotz}, Paul J. and {Lupton}, Robert H. and {Lust}, Nate B. and {MacArthur}, Lauren A. and {Mahabal}, Ashish and {Mandelbaum}, Rachel and {Markiewicz}, Thomas W. and {Marsh}, Darren S. and {Marshall}, Philip J. and {Marshall}, Stuart and {May}, Morgan and {McKercher}, Robert and {McQueen}, Michelle and {Meyers}, Joshua and {Migliore}, Myriam and {Miller}, Michelle and {Mills}, David J.},
        title = "{LSST: From Science Drivers to Reference Design and Anticipated Data Products}",
      journal = {\apj},
         year = 2019,
        month = mar,
       volume = {873},
       number = {2},
          eid = {111},
        pages = {111},
          doi = {10.3847/1538-4357/ab042c},
archivePrefix = {arXiv},
       eprint = {0805.2366},
 primaryClass = {astro-ph},
       adsurl = {https://ui.adsabs.harvard.edu/abs/2019ApJ...873..111I}
}

@ARTICLE{Green2019ApJ,
       author = {{Green}, Gregory M. and {Schlafly}, Edward and {Zucker}, Catherine and {Speagle}, Joshua S. and {Finkbeiner}, Douglas},
        title = "{A 3D Dust Map Based on Gaia, Pan-STARRS 1, and 2MASS}",
      journal = {\apj},
         year = 2019,
        month = dec,
       volume = {887},
       number = {1},
          eid = {93},
        pages = {93},
          doi = {10.3847/1538-4357/ab5362},
archivePrefix = {arXiv},
       eprint = {1905.02734},
 primaryClass = {astro-ph.GA},
       adsurl = {https://ui.adsabs.harvard.edu/abs/2019ApJ...887...93G}
}

@ARTICLE{Roman2020,
       author = {{Akeson}, Rachel and {Armus}, Lee and {Bachelet}, Etienne and {Bailey}, Vanessa and {Bartusek}, Lisa and {Bellini}, Andrea and {Benford}, Dominic and {Bennett}, David and {Bhattacharya}, Aparna and {Bohlin}, Ralph and et al.},
        title = "{The Wide Field Infrared Survey Telescope: 100 Hubbles for the 2020s}",
      journal = {arXiv e-prints},
         year = 2019,
        month = feb,
          eid = {arXiv:1902.05569},
        pages = {arXiv:1902.05569},
          doi = {10.48550/arXiv.1902.05569},
archivePrefix = {arXiv},
       eprint = {1902.05569},
 primaryClass = {astro-ph.IM},
       adsurl = {https://ui.adsabs.harvard.edu/abs/2019arXiv190205569A}
}

@ARTICLE{ZhanHu2011SSPMA,
       author = {{Zhan}, Hu},
        title = "{Consideration for a large-scale multi-color imaging and slitless spectroscopy survey on the Chinese space station and its application in dark energy research}",
      journal = {Scientia Sinica Physica, Mechanica \& Astronomica},
         year = 2011,
        month = jan,
       volume = {41},
       number = {12},
        pages = {1441},
          doi = {10.1360/132011-961},
       adsurl = {https://ui.adsabs.harvard.edu/abs/2011SSPMA..41.1441Z}
}

@ARTICLE{Gong2019ApJ,
       author = {{Gong}, Yan and {Liu}, Xiangkun and {Cao}, Ye and {Chen}, Xuelei and {Fan}, Zuhui and {Li}, Ran and {Li}, Xiao-Dong and {Li}, Zhigang and {Zhang}, Xin and {Zhan}, Hu},
        title = "{Cosmology from the Chinese Space Station Optical Survey (CSS-OS)}",
      journal = {\apj},
         year = 2019,
        month = oct,
       volume = {883},
       number = {2},
          eid = {203},
        pages = {203},
          doi = {10.3847/1538-4357/ab391e},
archivePrefix = {arXiv},
       eprint = {1901.04634},
 primaryClass = {astro-ph.CO},
       adsurl = {https://ui.adsabs.harvard.edu/abs/2019ApJ...883..203G}
}




\appendix

\section{Use of F96 response functions}

While system throughputs from D10 provide a more accurate characterization of the SDSS $ugriz$ photometric system, several early studies were based on those from F96. Here we use the latter to repeat the analysis presented in Section~\ref{sec:results} and \ref{sec:discus}. The most significant difference is a reduction in the $u$-band offset and improved consistency between the GALAH and APOGEE samples.

The SDSS photometric calibration is ultimately anchored to the spectrophotometric flux of $\rm{BD}+17^{\circ }4708$. Synthetic magnitudes of this star were computed using the F96 passbands, thereby defining the absolute scale of the system. Observations of BD+17°4708 with the USNO telescope were then used by \cite{smith02} to establish a network of 158 primary standards in the $u'g'r'i'z'$ (primed) system. The SDSS Photometric Telescope (PT) transferred this calibration to a set of secondary standard fields, which were subsequently used to calibrate imaging data obtained with the SDSS 2.5-m telescope in the $ugriz$ (unprimed) system.

In practice, the effective bandpasses differed slightly among the USNO, the PT, and the 2.5-m telescopes because the interference filters operate in different optical environments (in particular, the SDSS camera filters are located in vacuum). These differences lead to small shifts in effective wavelength and require colour-dependent transformations between the primed and unprimed systems. Additional complications arise from instrumental effects such as the $u$-band red leak. More accurate measurements of the total system throughput of the SDSS 2.5-m telescope, including filters, optics, and detector response are provided by D10 which we adopt as reference through this paper.

\begin{figure*}
    \includegraphics[scale=0.8]{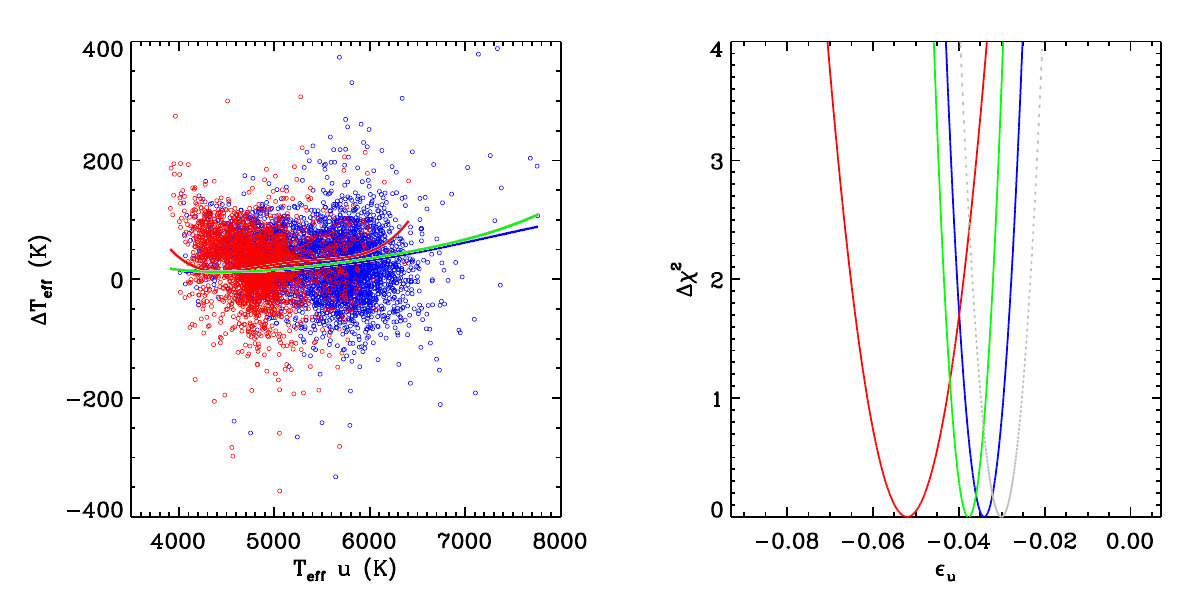}
    \caption{Same as Figure \ref{fig:epsilon_u}, but using F96.}
\end{figure*}

\begin{figure*}
    \includegraphics[scale=0.8]{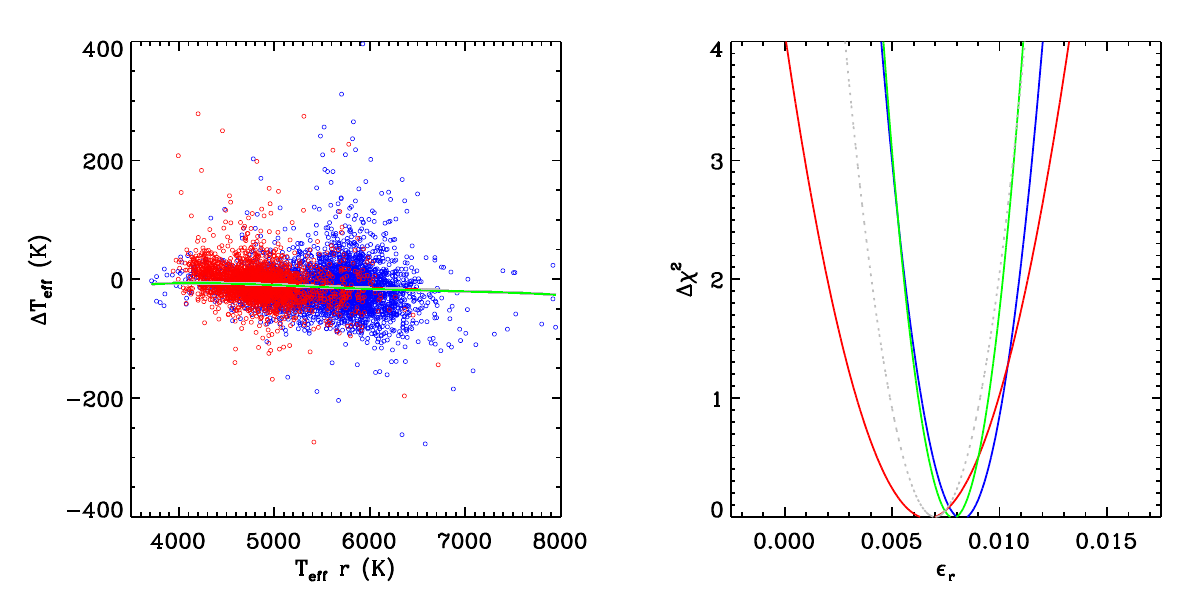}    
    \caption{Same as Figure \ref{fig:epsilon_r}, but using F96.}
\end{figure*}

\begin{table}
	\centering
	\caption{Same as Table \ref{tab:zp} (uncertainties omitted for clarity), but using responses from F96.}
	\begin{tabular}{lrrr} 
		\hline
		    & GALAH & APOGEE & COMBINED\\
		\hline
		$u$ & $-0.034$ & $-0.052$ & $-0.038$\\
		$g$ &  $0.026$ & $ 0.017$ & $ 0.024$ \\
		$r$ &  $0.008$ & $ 0.007$ & $ 0.008$\\
		$i$ &  $0.034$ & $ 0.048$ & $ 0.038$\\
		$z$ &  $0.045$ & $ 0.057$ & $ 0.048$\\        
		\hline
	\end{tabular}
\end{table}

\begin{figure*}\label{fig:calspec_F96}
    \includegraphics[scale=0.42]{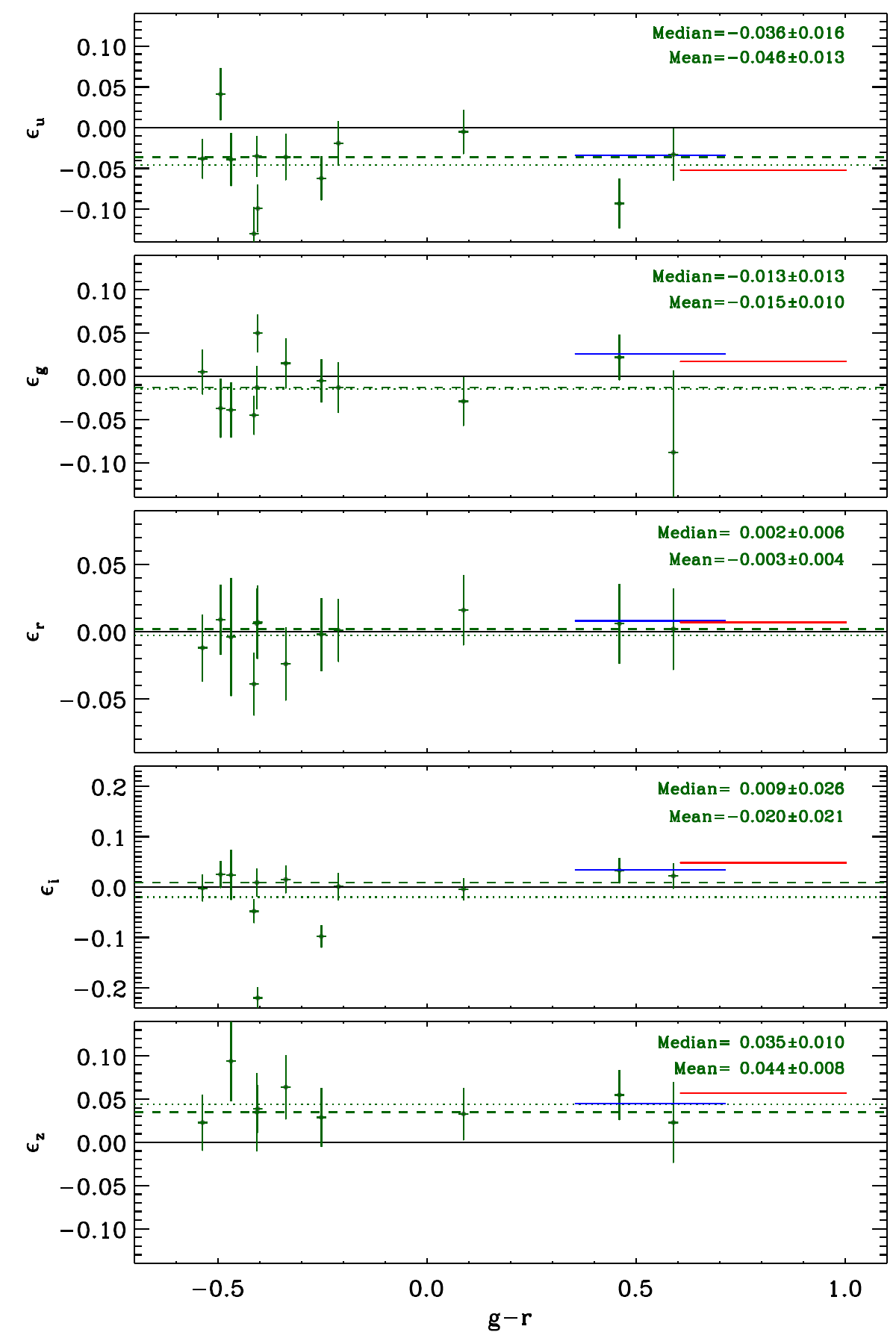}    
    \caption{Same as Figure \ref{fig:calspec_D10}, but using F96.}
\end{figure*}


\bsp	
\label{lastpage}
\end{document}